%% file: A2261Coe.tex
\documentclass[iop]{emulateapj}

\usepackage[colorlinks=true,linkcolor=blue,anchorcolor=blue,citecolor=blue,urlcolor=blue]{hyperref}  % allcolors=blue
\hypersetup{pdfauthor={Dan Coe}, pdftitle={CLASH: A2261 HST Strong Lensing Analysis}}
\newcommand\etal{et al.\ }

\newcommand\rmag{\ifmmode r_{625}\else$r_{625}$\fi}
\newcommand\imag{\ifmmode i_{775}\else$i_{775}$\fi}
\newcommand\zmag{\ifmmode z_{850}\else$z_{850}$\fi}
\newcommand\Vmag{\ifmmode V_{606}\else$V_{606}$\fi}
\newcommand\Imag{\ifmmode I_{814}\else$I_{814}$\fi}

\newcommand\rtwoh{\ifmmode {\rm r}_{200}\else r$_{200}$\fi}
\newcommand\ls{\mathrel{\hbox{\rlap{\hbox{\lower4pt\hbox{$\sim$}}}\hbox{$<$}}}}
\newcommand\gs{\mathrel{\hbox{\rlap{\hbox{\lower4pt\hbox{$\sim$}}}\hbox{$>$}}}}
\newcommand\BJ{$B_{\rm J}$}
\newcommand\VJ{$V_{\rm J}$}
\newcommand\RC{$R_{\rm C}$}
\newcommand\BVR{\BJ\VJ\RC}
\newcommand\VR{\VJ\RC}
\newcommand\ip{$i'$}
\newcommand\zp{$z'$}
\newcommand\BRz{\BJ\RC\zp}
\newcommand\BVRiz{\BJ\VJ\RC\ip\zp}
\newcommand\iz{\ip\zp}

\newcommand\VJm{V_{\rm J}}
\newcommand\RCm{R_{\rm C}}

\def\simgreat{\ifmmode{\mathrel{\mathpalette\@versim>}}
    \else{$\mathrel{\mathpalette\@versim>}$}\fi}
\def\simless{\ifmmode{\mathrel{\mathpalette\@versim<}}
    \else{$\mathrel{\mathpalette\@versim<}$}\fi}

\newcounter{thefigs}

\newcounter{thetabs}

\newcommand{\hGpc}{\ifmmode{h^{-1}{\rm Gpc}}\;\else${h^{-1}}${\rm Gpc}\fi}
\newcommand{\hkpc}{\ifmmode{h^{-1}_{70}\ {\rm kpc}}\;\else${h^{-1}_{70}}$ {\rm kpc}\fi}
\newcommand{\hMpc}{\ifmmode{h^{-1}_{70}\ {\rm Mpc}}\;\else${h^{-1}_{70}}$ {\rm Mpc}\fi}
\newcommand{\LCDM}{$\Lambda$CDM}

\def\simless{\mathbin{\lower 3pt\hbox
	{$\,\rlap{\raise 5pt\hbox{$\char'074$}}\mathchar"7218\,$}}} % < or of order
\def\simgreat{\mathbin{\lower 3pt\hbox
	{$\,\rlap{\raise 5pt\hbox{$\char'076$}}\mathchar"7218\,$}}} % > or of order

\newcommand{\HST}{{\em HST}}

\newcommand{\ACS}{{\em ACS}}
\newcommand{\WFCiii}{{\em WFC3}}

\newcommand{\hh}{h_{70}^{-1}}

\newcommand{\xfm}{\times 10^{15} M_\odot}
\newcommand{\xfmh}{\xfm \hh}

\newcommand{\Mvirsphd}{2.21^{+0.21}_{-0.15}}
\newcommand{\cvirsphd}{6.2^{+0.3}_{-0.3}}
\newcommand{\Mvirsph}{2.2\pm0.2}
\newcommand{\cvirsph}{6.2\pm0.3}
\newcommand{\Mres}{$M_{vir} = \Mvirsph \xfmh$}
\newcommand{\cres}{$c_{vir} = \cvirsph$}

\newcommand{\Mvirelld}{1.70^{+0.16}_{-0.12}}
\newcommand{\cvirelld}{4.6^{+0.2}_{-0.2}}
\newcommand{\Mvirell}{1.7\pm0.2}
\newcommand{\cvirell}{4.6\pm0.2}
\newcommand{\Mell}{$M_{vir} = \Mvirell \xfmh$}
\newcommand{\cell}{$c_{vir} = \cvirell$}

\shortauthors{Coe \etal 2011}
\slugcomment{Submitted to the Astrophysical Journal, \today}

\begin{document}

%\title{CLASH: Early Formation of Abell 2261 from Multiple Probes,
%Including Strong Lensing Analysis of 16-Band Hubble Imaging}
%\title{CLASH: Abell 2261 Not Over-Concentrated After All: 
%New Constraints from Strong Lensing Analysis of 16-Band Hubble Imaging}
%\title{CLASH: Abell 2261 Mass Concentration Uncertainties Resolved by Strong Lensing Analysis of 16-Band Hubble Imaging}
%\title{CLASH: Abell 2261 Mass Profile Restored to Agreement with Simulations by Strong Lensing Analysis of 16-Band Hubble Imaging}
%\title{CLASH: Precise New Measurements of the Mass Concentration of Abell 2261
%Based in Part on Strong Lensing Analysis of 16-Band Hubble Imaging}
%\title{CLASH: Precise New Constraints on the Mass Profile of Abell 2261
%from Strong Lensing Analysis of 16-Band Hubble Imaging}
%\title{CLASH: Precise Mass Profile of Abell 2261
%including New Strong Lensing Constraints from 16-Band Hubble Imaging}
\title{CLASH: Precise New Constraints on the Mass Profile of Abell 2261}
%from Strong Lensing Analyses of 16-Band Hubble Imaging}

%\shorttitle{CLASH: Abell 2261 Mass Profile}
\shorttitle{CLASH: Precise Mass Profile of Abell 2261}
%\shorttitle{CLASH: Abell 2261 HST Strong Lensing Analysis}
%\shorttitle{CLASH: Abell 2261 HST Strong Lensing Analysis}

%CLASH: New Constraints on the Inner Mass Profile of Abell 2261 Bring Its Concentration Back Into Agreement with Predictions from Simulations
%CLASH: New Constraints on the Inner Mass Profile of Abell 2261 Reveal a Concentration in Agreement with Predictions from Simulations
%CLASH: Abell 2261 Not Over-Concentrated After All: New Constraints from Strong Lensing Analysis of 16-Band HST Imaging
%New Evidence for 
%The Inner Mass Profile of Abell 2261 Constrained By Strong Lensing Analysis of 16-Band HST Imaging: Evidence for 

%Including Its First Strong Lensing Analysis from 16-Band Hubble Imaging}
%\title{CLASH: Another Galaxy Cluster with a Surprisingly Dense Concentrated Core and Strong Lensing Properties
%Revealed in 16-Band Hubble Imaging: Abell 2261}
% 16-band Hubble imaging
% Adi's suggestion -- CLASH: Joint strong+weak lensing analysis of Abell 2261 revealing another over-concentrated cluster

\author{Dan Coe\altaffilmark{1}}
% Contributors:
\author{Keiichi Umetsu\altaffilmark{2}}  % WL + lengthy discussions
\author{Adi Zitrin\altaffilmark{3}}  % SL
\author{Megan Donahue\altaffilmark{4}}  % X-ray
\author{Elinor Medezinski\altaffilmark{5}}  % Subaru, WL
\author{Marc Postman\altaffilmark{1}}  % BCG subtraction, detailed comments
\author{Mauricio Carrasco\altaffilmark{6,7}}  % SL
\author{Timo Anguita\altaffilmark{6,8}}  % SL
\author{Margaret J.~Geller\altaffilmark{9}}  % Hectospec
\author{Kenneth J.~Rines\altaffilmark{9,10}}  % Hectospec
\author{Antonaldo Diaferio\altaffilmark{11,12}}  % Hectospec
\author{Michael J.~Kurtz\altaffilmark{9}}  % Hectospec
\author{Larry Bradley\altaffilmark{1}}  % HST
\author{Anton Koekemoer\altaffilmark{1}}  % HST
\author{Wei Zheng\altaffilmark{5}}  % HST
\author{Mario Nonino\altaffilmark{13}}  % Subaru
\author{Alberto Molino\altaffilmark{14}}  % Subaru BPZ
\author{Andisheh Mahdavi\altaffilmark{15}}  % X-ray: JACO
\author{Doron Lemze\altaffilmark{5}}  % dynamical masses + comments
\author{Leopoldo Infante\altaffilmark{6}}  % with Mauricio & Timo
\author{Sara Ogaz\altaffilmark{1}}  % HST
\author{Peter Melchior\altaffilmark{16}}  % discussion: WL
\author{Ole Host\altaffilmark{17}}  % discussion incl. axis ratio
\author{Holland Ford\altaffilmark{5}}  % discussion and comments
\author{Claudio Grillo\altaffilmark{18}}  % BCG vel disp. + dicsussion
\author{Piero Rosati\altaffilmark{7}}  % comments
\author{Yolanda Jim\'enez-Teja\altaffilmark{14}}  % BCG subtraction
\author{John Moustakas\altaffilmark{19}}  % stellar profile
\author{Tom Broadhurst\altaffilmark{20}}  % discussion c-M
\author{Bego\~na Ascaso\altaffilmark{14}}  % discussion cluster evolution
\author{Ofer Lahav\altaffilmark{17}}  % discussion
%% other CLASH core members - in alphabetical order:
\author{Matthias Bartelmann\altaffilmark{3}} 
\author{Narciso Ben\'itez\altaffilmark{14}}
\author{Rychard Bouwens\altaffilmark{21}}
\author{Or Graur\altaffilmark{22}}
\author{Genevieve Graves\altaffilmark{23}}
\author{Saurabh Jha\altaffilmark{24}}
\author{Stephanie Jouvel\altaffilmark{17}}
\author{Daniel Kelson\altaffilmark{25}}
\author{Leonidas Moustakas\altaffilmark{26}}
\author{Dan Maoz\altaffilmark{22}}
\author{Massimo Meneghetti\altaffilmark{27}}
\author{Julian Merten\altaffilmark{26}}
\author{Adam Riess\altaffilmark{1,5}}
\author{Steve Rodney\altaffilmark{5}}
\author{Stella Seitz\altaffilmark{28}}

\email{DCoe@STScI.edu}

\altaffiltext{1}{Space Telescope Science Institute, 3700 San Martin Drive, Baltimore, MD 21218, U.S.A.}
\altaffiltext{2}{Academia Sinica, Institute of Astronomy \& Astrophysics}
\altaffiltext{3}{Universitat Heidelberg}
\altaffiltext{4}{Michigan State University}
\altaffiltext{5}{Johns Hopkins University}
%\altaffiltext{6}{Universidad Cat\'olica de Chile}
\altaffiltext{6}{AIUC-DAA, Pontificia Universidad Cat\'olica de Chile}
\altaffiltext{7}{European Southern Observatory}
\altaffiltext{8}{Max Planck Institut f\"ur Astronomie, Heidelberg}
\altaffiltext{9}{Smithsonian Astrophysical Observatory}
\altaffiltext{10}{Western Washington University}
% Western Washington University, Department of Physics and Astronomy, Bellingham, WA 98225
\altaffiltext{11}{Universit\`a degli Studi di Torino}
\altaffiltext{12}{INFN, Sezione di Torino}
\altaffiltext{13}{INAF, Osservatorio Astronomico di Trieste}
\altaffiltext{14}{Instituto de Astrof\'isica de Andaluc\'ia}
\altaffiltext{15}{San Francisco State University} % Dept of Physics and Astronomy, Thornton Hall 527, amahdavi@sfsu.edu
\altaffiltext{16}{Ohio State University}
\altaffiltext{17}{University College London}
\altaffiltext{18}{Excellence Cluster Universe, Technische Universit\"at M\"unchen}
%\altaffiltext{19}{Center for Astrophysics and Space Sciences, University of California, San Diego}
\altaffiltext{19}{University of California, San Diego}
\altaffiltext{20}{University of the Basque Country}
%\altaffiltext{20}{Department of Theoretical Physics, University of the Basque Country}
\altaffiltext{21}{University of Leiden}
\altaffiltext{22}{Tel Aviv University}
\altaffiltext{23}{University of California at Berkeley}
\altaffiltext{24}{Rutgers University}
\altaffiltext{25}{Carnegie Institute for Science; Carnegie Observatories}
\altaffiltext{26}{Jet Propulsion Laboratory, California Institute of Technology}
\altaffiltext{27}{INAF, Osservatorio Astronomico di Bologna}
\altaffiltext{28}{Universitas Sternwarte, M\"unchen}
%3 Dipartimento di Fisica Generale ÒAmedeo Avogadro,Ó Universit`a degli Studi di Torino, Torino, Italy
%4 Istituto Nazionale di Fisica Nucleare (INFN), Sezione di Torino, Torino, Italy

%amahdavi@sfsu.edu

%\author{T. Anguita}
%\affil{Centro de Astro-Ingenier\'ia, Departamento de Astronom\'ia y Astrof\'isica, Pontificia Universidad Cat\'olica de Chile, Casilla 306, Santiago, Chile.}
%\affil{Max-Planck-Institut f\"ur Astronomie, K\"onigstuhl 17, 69117 Heidelberg, Germany.}

\begin{abstract}

We precisely constrain the inner mass profile of Abell 2261 ($z = 0.225$) for the first time
and determine this cluster is not ``over-concentrated'' as found previously,
implying a formation time in agreement with \LCDM\ expectations.
These results are based on strong lensing analyses of new 16-band HST imaging
obtained as part of the Cluster Lensing and Supernova survey with Hubble (CLASH).
Combining this
%with revised weak lensing analyses of 5-band Subaru + KPNO wide field imaging,
with revised weak lensing analyses of Subaru wide field imaging
with 5-band Subaru + KPNO photometry,
we place tight new constraints on the halo 
virial mass \Mres\ (within $r_{vir} \approx 3$ Mpc $\hh$)
and concentration \cres\ when assuming a spherical halo.
%reveals that A2261 is 
%not in fact ``over-concentrated'' as found previously but rather 
%This is in good agreement with $c(M,z)$ predictions from \LCDM\ simulations.
This agrees broadly with average $c(M,z)$ predictions from recent \LCDM\ simulations
which span $5 \lesssim \langle c \rangle \lesssim 8$.
%$\langle c \rangle \sim 5--8$.
%
Our most significant systematic uncertainty is halo elongation along the line of sight.
%To estimate possible halo elongation along the line of sight,
To estimate this,
we also derive a mass profile based on archival Chandra X-ray observations
and find it to be $\sim 35\%$ lower than our lensing-derived profile at $r_{2500} \sim 600$ kpc.
%(with better agreement in the inner core).
%
Agreement 
%between our lensing and X-ray mass profiles
can be achieved by a halo elongated with a $\sim 2:1$ axis ratio along our line of sight.
%with increased sphericity inside $r \lesssim 100$ kpc
%(roughly the visible extent of the BCG).
%
For this elongated halo model, we find
\Mell\ and \cell,
%$M_{vir} = 1.65^{+0.16}_{-0.12} \xfmh$
%and $c_{vir} = 4.7^{+0.2}_{-0.3}$,
placing rough lower limits on these values.
%
%This lower concentration
%$c_{vir} \sim 4.9$ 
%is still in agreement with many simulations 
%but lower than one of the more recent predictions.
%for a massive $M_{vir} \sim 2 \times 10^{15} M_{\odot}$ cluster at $z \sim 0.2$.
%
The need for halo elongation can be partially obviated by non-thermal pressure support 
and, perhaps entirely, by systematic errors in the X-ray mass measurements.
%as evidenced by one published result that is in excellent agreement with our lensing-derived spherical mass profile.
%
We estimate the effect of background structures based on MMT/Hectospec spectroscopic redshifts
and find these tend to lower $M_{vir}$ further by $\sim 7\%$ and increase $c_{vir}$ by $\sim 5\%$.
%
%We note that published dynamical mass estimates of A2261 are significantly lower, 
%but these data are somewhat limited by bright stars in this field,
%hindering our ability to obtain additional spectra which might resolve this discrepancy.
%hindering our ability to resolve this discrepancy.
%
%Finally we discuss quantitative estimates of the formation time of A2261
%based on the observed concentration and other observational probes.

\end{abstract}


\section{Introduction}
\label{sec:intro}

Detailed observational constraints of dark matter halos yield
important tests to our understanding of structure formation \citep{Natarajan07,BorganiKravtsov11},
%cosmology \citep{DETF09},
the particle nature of dark matter \citep{Clowe06,KeetonMoustakas09},
and perhaps the nature of dark energy as well \citep{GrossiSpringel09}.
Large cluster surveys require precisely determined cluster masses to calibrate their observables
and achieve their full potential to constrain cosmology
\citep{Henry09,Allen11}.
%\citep{Voit05,Henry09,Allen11}.  % cause Henry09 to wrap around line and give pdftex error!
%\citep[see recent review by][]{Allen11}.

%The best studied galaxy clusters appear to have 
The galaxy clusters studied best via gravitational lensing
appear to have more densely concentrated cores
than clusters of similar mass and redshift formed in \LCDM\ simulations
\citep{Broadhurst08,BroadhurstBarkana08,Oguri09,Richard10,Sereno10,Zitrin11MACS,CLASH}.
(See results from other methods reviewed in \citealt{Fedeli11} and \citealt{Bhattacharya12}.)
%(\citealt{Broadhurst08,BroadhurstBarkana08,Oguri09,Richard10,Sereno10,Zitrin11MACS,CLASH};
%see results from other methods reviewed in \citealt{Fedeli11}).
%\citep{Broadhurst08,BroadhurstBarkana08,Oguri09,Richard10,Sereno10,Zitrin11MACS,CLASH}.
%
Some of this discrepancy is due to bias,
as the clusters selected for the most detailed lensing studies
were among the strongest gravitational lenses known.
However it is estimated that even this large ($\sim 50$ -- 100\%) bias 
cannot fully explain the high observed concentrations
(\citealt{Hennawi07,Oguri09,Meneghetti10,Meneghetti11}, although see \citealt{Oguri11}).
Baryons, absent from these dark matter only simulations,
are found to only modify cluster concentrations at the $\lesssim 10\%$ level \citep{Duffy10,Mead10,Fedeli11}.
%See \cite{Fedeli11} for a recent review of observational results from various methods.

If confirmed, this result would imply
that galaxy clusters formed earlier than their counterparts in simulated \LCDM\ universes.
We expect that the higher density of the earlier universe
would remain imprinted on the cluster cores
as we observe them today
\citep[e.g.,][]{Jing00, Bullock01,Wechsler02, Zhao03a}.

Another possible hint of early cluster formation
may be galaxy clusters detected at $z > 1$
which are perhaps unexpectedly massive
%Fassbender11,Nastasi11
\citep{Stanford06,Eisenhardt08,Jee09,Huang09,Rosati09,Papovich10,Schwope10,Gobat11,Jee11,Foley11,
Santos11a,Santos11b,Planck11clus}.
However one should be cautious about the statistical interpretation of such results 
\citep{Paranjape11,Hotchkiss11,Hoyle11,Waizmann11,HarrisonColes11,HarrisonColes11b}.
Building on results from large X-ray surveys \citep{Vikhlinin09c,Mantz10},
%Clusters detected in 
large new Sunyaev-Zel'dovich (SZ) surveys will continue to 
%yield interesting cosmological constraints
constrain cosmology
based on cluster abundance measurements as functions of mass and redshift
\citep{Sehgal11,Benson11}.
%provide interesting constraints on observed numbers of high redshift clusters
%\citep{Sehgal11,Williamson11}.

%Unexpectedly early formation times have also been suggested for individual galaxies 
%discovered at high redshift $z \gtrsim 6$ 
%(e.g., \citealt{Richard11}, although see \citealt{Pirzkal11}).
%Such results are made possible in part by cluster lensing magnification
%which enables more detailed study of these distant objects than otherwise possible.
%\citep[e.g.,][]{Bradley08,Hall11,Zitrin11MACS0329}.
%Bouwens09,Maizy10 -- overall counts

%Mechanisms proposed to explain such early growth on both galaxy and cluster scales
Mechanisms proposed to explain such early growth
include departures from an initially Gaussian spectrum of density fluctuations
\citep[e.g.,][]{ChongchitnanSilk11},
though we note some such non-Gaussian models 
can be ruled out based on cosmic X-ray background measurements \citep{Lemze09b}.
Early growth may also be explained by higher levels of dark energy in the past.
This idea, dubbed Early Dark Energy (EDE; \citealt{FedeliBartelmann07,SadehRephaeli08,Francis09,GrossiSpringel09}),
would have suppressed structure growth in the early universe,
such that clusters would have had to start forming sooner to yield the numbers we observe today.
Other dark energy theories with similar implications have also been proposed
\citep[e.g.,][]{Baldi11,Carlesi11}.
%Other ideas with similar implications have been proposed,
%including ``bouncing coupled dark energy'' \citep{Baldi11}
%and ``vector dark energy'' \citep{Carlesi11}

Significant improvements in these observational constraints are being obtained by
CLASH, the Cluster Lensing and Supernova survey with Hubble
\citep{CLASH}.\footnote{\href{http://www.stsci.edu/\%7Epostman/CLASH/}{http://www.stsci.edu/$\sim$postman/CLASH}}
CLASH is a 524-orbit multi-cycle treasury \HST\ program
to observe 25 galaxy clusters ($0.18 < z < 0.89$) each in 16 filters
with the Wide Field Camera 3 (\WFCiii; \citealt{WFC3})
and the Advanced Camera for Surveys (\ACS; \citealt{ACS})
over the course of three years (\HST\ cycles 18-20).
Importantly, 20 CLASH clusters were X-ray selected to be massive and relatively relaxed.
This avoids the strong bias toward high concentrations
in previously well-studied clusters selected for their lensing strength.
%when clusters are selected for study based on strong lensing strength,
%as was the case for the previously best studied clusters.

%Abell 2261 (hereafter, A2261) was the third CLASH cluster to be observed and the second of our X-ray selected clusters.
%Abell 2261 (hereafter, A2261) was the second of our X-ray selected clusters to be observed.
Abell 2261 (hereafter, A2261) was observed as part of the CLASH program.
%It has a redshift of $z = 0.2249$ \citep{Crawford95,Rines10}.  % Ebeling96
It has a redshift of $z = 0.2249$ as measured by \cite{Crawford95} and refined by \cite{Rines10}.  % Ebeling96
%It has a redshift of $z = 0.2242$ as measured by \cite{Crawford95} and \cite{Rines10}
%and refined most recently by XX et al.~in preparation.  % Ebeling96

Weak lensing (WL) analyses of ground-based imaging of A2261 \citep{Umetsu09,Okabe10}
yielded concentration measurements of $c_{vir} \sim 6$ or $\sim 10$,
with the broad range attributed to measurement uncertainties, the details of the analysis method used,
and perhaps subject to uncertainty due to massive background structures identified at $z \sim 0.5$.
The latter value ($c_{vir} \sim 10$) would be
significantly higher than predicted for an average relaxed cluster of A2261's mass and redshift:
$c_{vir} \sim 5$ from \cite{Duffy08},
although an analysis of more recent simulations 
yields a much higher prediction: $c_{vir} \sim 8.5$ \citep{Prada11}.
%$c_{vir} \sim 4$ or $\sim 7$ from simulations by \cite{Duffy08} and \cite{Prada11}, respectively.
The WL measurements had overlapping uncertainties,
but a preliminary strong lensing (SL) measurement of the Einstein radius 
($R_{E} \approx 40\arcsec$ for a background source at $z_{s} = 1.5$)
supported the larger value with smaller uncertainties:
$c_{vir} = 11 \pm 2$ 
%$c_{vir} = 11.1^{+2.2}_{-1.9}$ 
\citep{Umetsu09}.
This result was also included in \cite{Oguri09}
as one of ten well-studied clusters,
all of which had higher than predicted concentrations.

In this work, we revisit both the strong and weak lens modeling.
Our deep 16-band HST imaging
reveals strongly lensed (multiply imaged) galaxies 
all undetected in the previous HST imaging 
%(0.5-orbit \WFPCii\ F606W)
(0.5-orbit WFPC2 F606W)
and allows us to derive robust and precise photometric redshifts for these arcs,
a key ingredient for our mass model of the cluster core.

Detailed strong lensing analysis is required to 
accurately and precisely measure the inner mass profile and concentration of A2261.
By probing the mass profile over a combined two decades of radius,
joint analysis of strong plus weak lensing yields significantly higher precision measurements
of cluster virial masses and concentrations than either method alone
%as demonstrated quantitatively by 
\citep{Meneghetti10}.
%While both are required,
%weak lensing more strongly constrains the outer profile and virial mass,
%while strong lensing better constrains the inner profile and the mass concentration.

%OTHER BACKGROUND ON A2261.  PREVIOUS WORK.
%XX: OTHER PREVIOUS NON-LENSING WORK ON A2261...

This paper is organized as follows.
We describe our HST (\S\ref{sec:HST}) and MMT spectroscopic (\S\ref{sec:specz}) observations
followed by our strong lens mass modeling (\S\ref{sec:SL}).
We then introduce our ground-based imaging and weak lensing analyses (\S\ref{sec:WL})
and derive joint strong + weak lensing constraints (\S\ref{sec:lensing}).
We constrain halo triaxiality from joint lensing + X-ray constraints in \S\ref{sec:joint}
and finally compare our mass profile with results from simulations in \S\ref{sec:c-M}.
The formation time of A2261 is discussed in a broader context
including other observational probes in \S\ref{sec:probes},
and we summarize our conclusions in \S\ref{sec:conclusions}.

%\citep{Sand05} -- single arc in 0.5 orbit F606W WFPC2

%\citep{Rines10} mass estimates

Where necessary to calculate distances, etc.,
we assume a concordance \LCDM\ cosmology
%with $h = 0.7 h_{70}$, $\Omega_m = 0.3$, $\Omega_\Lambda = 0.7$,
%where $H_0 = 100$ $h$ km s$^{-1}$ Mpc$^{-1}$ and $h_{70} \approx 1$ (which we will often omit).
with $h = 0.7$, $\Omega_m = 0.3$, $\Omega_\Lambda = 0.7$,
%($h_{70} = 1$)
where $H_0 = 100$ $h$ km s$^{-1}$ Mpc$^{-1}$.
%and $h_{70} = h / 0.7 \approx 1$ (which we will often omit).
%In this cosmology, $1\arcsec \approx 3.59$ kpc $h_{70}^{-1} \approx 2.51$ kpc $h^{-1}$
%at A2261's redshift of $z = 0.224$,
%where we define $h = 0.7 h_{70}$.
In this cosmology,
at A2261's redshift of $z = 0.225$,
$1\arcsec \approx 3.59$ kpc $h_{70}^{-1} \approx 2.51$ kpc $h^{-1}$,
where $h = 0.7 h_{70}$.
Furthermore at this redshift,
the cluster virial radius is defined as that which contains 
an average overdensity of $\Delta_c \approx 115$ times critical,
where $\Delta_c \approx 18 \pi^2 - 82 \Omega_\Lambda(z) - 39 \Omega^2_\Lambda(z)$
based on spherical collapse theory
\citep{BryanNorman98}.
%(Appendix \ref{sec:appendix}).

%Furthermore, a halo virial radius at $z = 0.224$ is defined as that radius which contains 
%an average overdensity of $\Delta_c \sim 115$ above the critical density necessary to close the universe,
%$\rho_{crit} = 3 H^2(z) / (8 \pi G)$
%\citep{BryanNorman98}.
%We refer the reader to the Appendix \ref{sec:appendix} for definitions of the virial mass and concentration.

%with $h = 0.7$, $\Omega_m = 0.3$, $\Omega_\Lambda = 0.7$,
%where $H_0 = 100$ $h$ km s$^{-1}$ Mpc$^{-1}$.
%In this cosmology, $1\arcsec \approx 3.59$ kpc $\approx 2.51$ kpc $h^{-1}$

%Furthermore, a halo virial radius at $z = 0.224$ is defined as that radius which contains 
%an average overdensity of $\Delta_c \sim 115$ above the critical density necessary to close the universe,
%$\rho_{crit} = 3 H^2(z) / (8 \pi G)$
%\citep{BryanNorman98}.
%We refer the reader to the Appendix \ref{sec:appendix} for definitions of the virial mass and concentration.

%We note that $M_{100}$ is evaluated 
%at a slightly larger radius than $M_{vir} = M_{115}$, 
%and thus should be a few percent higher,
%assuming roughly NFW profiles with concentrations as above \citep{Coe10DMprofiles}.

%%%%%%%%%%%%%%%%%%%%%%%%%%%%%%%%%
\begin{deluxetable}{cccc}
\tablewidth{0pt}
\tablecaption{\label{tab:obs}CLASH HST Observations of the Core of A2261}
\tablehead{
\colhead{Camera /}&
\colhead{Filter}&
\colhead{HST}&
\colhead{Exposure Time}\\
\colhead{Channel}&
\colhead{Element}&
\colhead{Orbits}&
\colhead{(sec)}
}
% ~/CLASH/data/a2261/amk/20110601/summary.txt
\startdata
WFC3/UVIS & F225W & 1.5 & 3671\\
WFC3/UVIS & F275W & 1.5 & 3745\\
WFC3/UVIS & F336W & 1.0 & 2408\\
WFC3/UVIS & F390W & 1.0 & 2456\\
ACS/WFC    & F435W & 1.0 & 2077\\
ACS/WFC    & F475W & 1.0 & 2064\\
ACS/WFC    & F606W & 1.0 & 2057\\
ACS/WFC    & F625W & 1.0 & 2064\\
ACS/WFC    & F775W & 1.0 & 2072\\
ACS/WFC    & F814W & 2.0 & 4099\\
ACS/WFC    & F850LP & 2.0 & 4148\\
WFC3/IR      & F105W & 1.0 & 2814\\
WFC3/IR      & F110W & 1.0 & 2514\\
WFC3/IR      & F125W & 1.0 & 2514\\
WFC3/IR      & F140W & 1.0 & 2411\\
WFC3/IR      & F160W & 2.0 & 5029\\
\vspace{-0.1in}
\enddata
\tablecomments{Parallel observations and supernova follow-up observations are not described here nor utilized in this work.}
%\tablenotetext{a}{Exposure times are the average time for each filter for all CLASH cycle 18 observations.}
\end{deluxetable}
%%%%%%%%%%%%%%%%%%%%%%%%%%%%%%%%%

\section{HST Observations}
\label{sec:HST}

%Prior to CLASH, A2261 (R.A. $= 17^{\rm h} 22^{\rm m} 27\fs2$, decl. $= +32\arcdeg 07\arcmin 57.3\arcsec$ [J2000])
%was observed with \HST\ only once back in \HST\ Cycle 8 (1999): 0.5-orbit (1000 seconds) in F606W with \WFPCii\ (GO 8301; PI Edge).
% http://archive.stsci.edu/proposal_search.php?mission=hst&id=8301

We observed 
A2261 (BCG R.A. $= 17^{\rm h} 22^{\rm m} 27\fs2$, decl. $= +32\arcdeg 07\arcmin 57\arcsec$ [J2000])
%A2261 (R.A. $= 17^{\rm h} 22^{\rm m} 27\fs2$, decl. $= +32\arcdeg 07\arcmin 57.3\arcsec$ [J2000])
as part of the CLASH program
in \HST\ Cycle 18 between 2011 Mar 9 and May 21 
to a total depth of 20 orbits
in 16 \WFCiii\ and \ACS\ filters, 
spanning $\sim$ 2,000\AA\ -- 17,000\AA\ (Table \ref{tab:obs}; GO 12066; PI Postman).
%
%\subsection{Image Processing}
%
The images were processed for debias, flats, superflats, and darks using standard techniques, 
and then co-aligned and combined using drizzle algorithms.
% to a scale of 0.065"/pixel.
See \cite{Koekemoer07} and \cite{CLASH} for details.

%\subsection{BCG modeling and subtraction}
\label{sec:galsub}

In order to better reveal faint lensed images,
we modeled and subtracted the BCG light in all 12 ACS+IR filters.
We used the isophote fitting routine, SNUC, which is part of the XVISTA image processing system, 
to derive two-dimensional models of the bright early-type galaxies in A2261, including the BCG. 
SNUC is capable of simultaneously obtaining the best non-linear least-squares fits 
to the two-dimensional surface brightness distributions
in multiple, overlapping galaxies \citep{Lauer86}. 
The models were derived independently for each CLASH passband. 
Fits were performed using concentric isophotes 
but the position angles and the ellipticities of the isophotes were left as free parameters. 
The models were then subtracted from the original image to produce a bright-galaxy subtracted image.

%\subsection{Galaxy detection, photometry, and photometric redshifts}
%\subsection{Photometry and photometric redshifts}
\label{sec:phot}

We used SExtractor \citep{SExtractor} to detect objects and measure their photometry.
For arcs eluding this initial detection, we constructed manual apertures
which were then forced back into SExtractor using SExSeg \citep{Coe06}.
%To correct our photometry for the range of observed PSFs
%(with FWHM's ranging from $\sim 0.07\arcsec$ to $\sim 0.15\arcsec$; \citealt{WFC3handbook}),
%aperture corrections are estimated as a function of wavelength and radius
%using published tables of observed encircled energy \citep{WFC3handbook,Sirianni03}.
Isophotal apertures were used as they have been shown to yield robust colors \citep{Benitez04}.

Based on this photometry, we measured photometric redshifts using BPZ \citep{Benitez00,Benitez04,Coe06}.
Spectral energy distribution (SED) templates are redshifted and fit to the observed photometry.
A Bayesian analysis tempers the qualities of fit with a prior:
the empirical likelihood of redshift as a function of both galaxy magnitude and type
(e.g., bright and/or elliptical galaxies are rare at high redshift).
Here we used 11 SED templates originally from PEGASE \citep{Fioc97}
but strongly recalibrated based on photometry and spectroscopic redshifts 
of galaxies in the FIREWORKS catalog \citet{Wuyts08}.
These templates yield $\lesssim 1\%$ outliers for high quality spectroscopic samples
and therefore implicitly encompass the full range of 
metallicities, extinctions, and star formation histories of real galaxies.

%Table \ref{tab:arcs}.
%Figure \ref{fig:arcs}.
%Figure \ref{fig:arcstamps}.

%%%%%%%%%%%%%%%%%%%%%%%%%%%%%%%%%
\begin{figure*}
\epsscale{1.15}
%\plotone{figs/arcslabeled.png}
%\plotone{figs/arcslabeled.jpg}
%\plottwo{figs/arcs/critcurvescrop.png}{figs/arcs/critcurvessubtcrop.png}
%\plottwo{figs/arcs/critcurves.png}{figs/arcs/critcurvessubt.png}
%\plottwo{figs/arcs/critcurves2.png}{figs/arcs/critcurvessubt2.png}
\plottwo{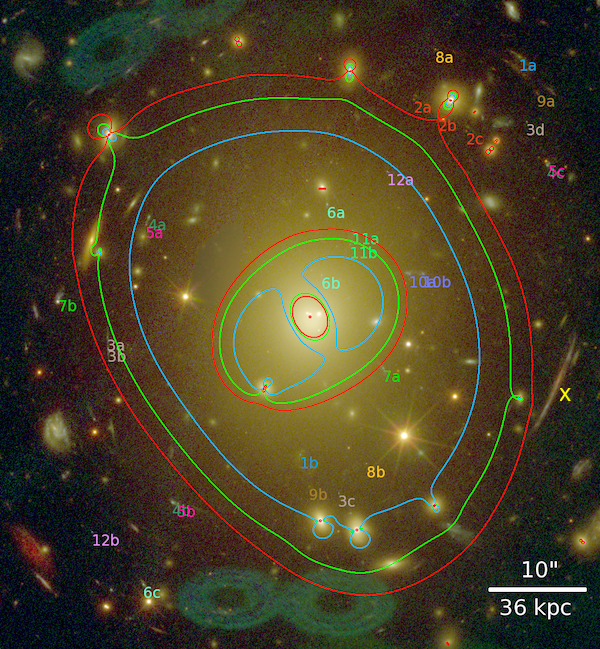}{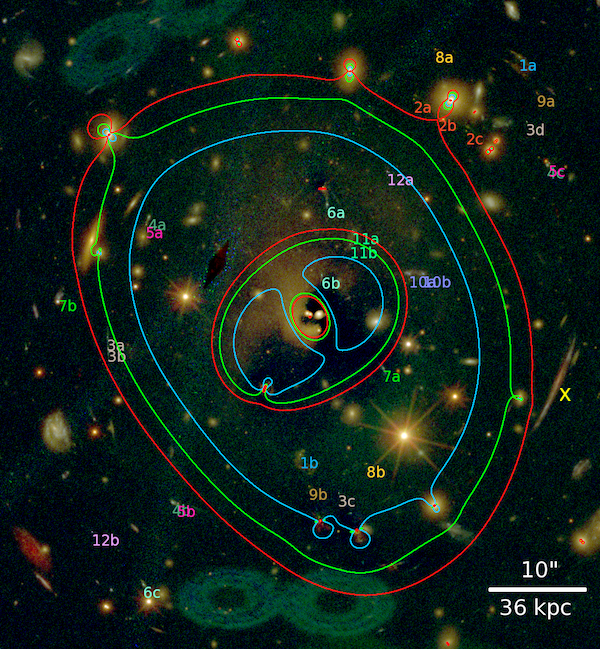}
\caption{\label{fig:arcs}
Multiple images of background galaxies strongly lensed by A2261.
All were identified in this work based on our deep, multiband \HST\ imaging and lens modeling.
Each is located directly above and to the left of its label.
Close ups may be seen in Fig.~\ref{fig:arcstamps}.
The prominent arc marked ``X'' is not multiply-imaged according to our models.
Lensing critical curves from our primary lens model are plotted for background sources 
at redshifts $z_s = 1$ (blue), 2 (green), and 7 (red).
%Each side of the image spans $XX \arcmin = YY$ kpc $h^{-1}$.
%The image spans $\sim 74\arcsec \times 69 \arcsec \approx 264 \times 246$ kpc.
%Each image is $65 \arcsec \sim 233$ kpc on a side.
These HST ACS+WFC3/IR color images were produced automatically using 
\href{http://www.stsci.edu/\%7Edcoe/trilogy/}{Trilogy} (\S\ref{sec:colorimages}) 
which reveals faint features without saturating bright areas such as the BCG core.
Filters were assigned colors as follows: 
Blue = F435W + F475W; Green = remaining ACS filters; Red = WFC3/IR.
%\footnote{\href{http://www.stsci.edu/\%7Edcoe/trilogy/}{http://www.stsci.edu/$\sim$dcoe/trilogy/}}.
%with filters as described in Table \ref{tab:colorfilters}.
%Filters were assigned colors as described in Table \ref{tab:colorfilters}.
The green ``figure 8'' patterns are ACS reflection artifacts from a bright star to the SE.
North is up, East is left.
\emph{Left:} The diamond-shaped hole in the ACS images was filled in with an IR image 
tinted yellow to roughly match the color of this portion of the BCG's stellar halo.
\emph{Right:} Light has been modeled and subtracted 
from the BCG and a few other cluster galaxies close to the arcs (\S\ref{sec:galsub}).
%The ``yin yang'' pattern revealed at the center of the BCG may be real structure or perhaps a modeling artifact 
%and will be investigated further by Postman et al.~in prep.
The residual pattern near the location of the BCG center 
(aside from the bright knots)
is most likely due to a combination of 
small model artifacts and real asymmetries in the stellar distribution. 
%The amplitude of the residual pattern is $<XXX \%$ of the actual BCG light within the central YY arcseconds.
}\end{figure*}
%%%%%%%%%%%%%%%%%%%%%%%%%%%%%%%%%

%\input{/Users/dcoe/A2261/SL/table/tabarcs}
%\input{figs/tabarcs}
\input{figs/arcs/tabarcs_patched}
% \ref{tab:arcs}

%%%%%%%%%%%%%%%%%%%%%%%%%%%%%%%%%
\begin{figure*}
\epsscale{1.17}
%\plotone{figs/arcs.png}
\plotone{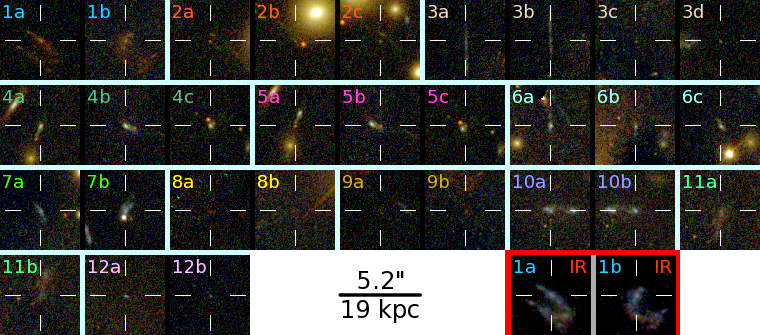}
\caption{\label{fig:arcstamps}
Multiple images of galaxies strongly lensed by A2261,
as identified in this work (see \S\ref{sec:SLimages}, Figure \ref{fig:arcs}, and Table \ref{tab:arcs}).
The BCG has been modeled and subtracted from all of these images.
%Most are ACS color images, while those at bottom right use WFC3/IR filters (see Table \ref{tab:colorfilters}).
Most are ACS color images 
(B = F435W+F475W, G = F606W+F625W, R = F775W+F814W+F850LP), 
while those at bottom right use WFC3/IR filters
(B = F105W+F110W, G = F125W+F140W, R = F160W).
Each image stamp is $5.2\arcsec \approx 19$ kpc on a side. % 80 pixels
%Color images use filters as described in Table \ref{tab:colorfilters}.
}\end{figure*}
%%%%%%%%%%%%%%%%%%%%%%%%%%%%%%%%%

%\subsection{Color Images}
\label{sec:colorimages}

The color images used in this paper were produced automatically using the publicly available 
Trilogy software.\footnote{\href{http://www.stsci.edu/\%7Edcoe/trilogy/}{http://www.stsci.edu/$\sim$dcoe/trilogy/}}
Trilogy determines the intensity scaling automatically
and independently in each color channel
to display faint features without saturating bright features.
The scalings are determined 
based on a sample of the summed images and two input parameters.
One sets the output luminosity of ``the noise'',
currently determined as 1-$\sigma$ above the sigma-clipped mean.
The other parameter sets what fraction of the data (if any) in the sample region should be allowed to saturate.
Default values for these parameters (0.15 and 0.001\%, respectively)
work well, but the user is able to tweak them.
The scaling is accomplished using the logarithmic function $y = a \log_{10} (k x + 1)$
clipped between 0 and 1,
where $a$ and $k$ are constants determined based on the data and desired scaling parameters as described above.
%The scaling is accomplished using the logarithmic function $y = \log_{10} (k x + 1) / r$,
%where the constants $k$ and $r$ are determined based on the data and desired scaling.
%The scaling is accomplished using the logarithmic function $y = \log_{10} (k (x - x_0) + 1) / r$,
%where the constants $k$, $x_0$, and $r$ are determined based on the data and desired scaling.
%We used various filter combinations to produce the color images shown in this work.
%These are given in Table \ref{tab:colorfilters}
%for our ACS, IR, and ACS+IR color images.  % UVIS

%%%%%%%%%%%%%%%%%%%%%%%%%%%%%%%%%
\begin{figure*}
\epsscale{1.17}
%\plotone{figs/sedcompare.png}
\plotone{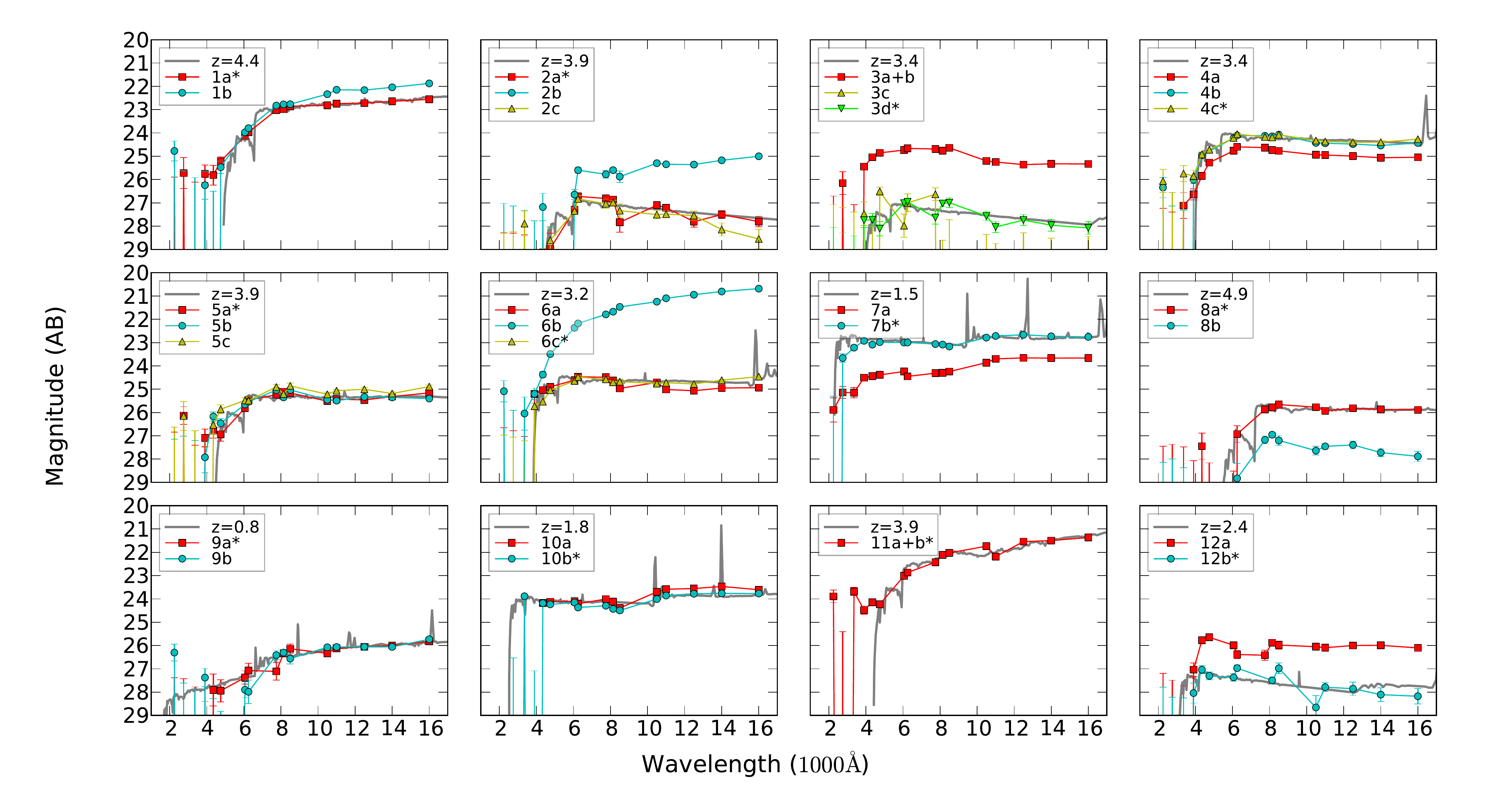}
\caption{\label{fig:sedcompare}
Spectral energy distributions (SEDs) of arcs as observed in our 16 HST filters.
Also plotted in gray are BPZ SED fits to the best isolated arcs marked with asterisks (*)
both in this figure and in Table \ref{tab:arcs}.
For each arc, the SEDs of all images should have similar shapes
though these may shift vertically in magnitude as their magnifications may vary.
The photometry of a few faint images may be contaminated by cluster light,
despite our best efforts to model and subtract the BCG and other cluster galaxies.
Most notably, 6b is a faint image very near the BCG core.
}\end{figure*}
%%%%%%%%%%%%%%%%%%%%%%%%%%%%%%%%%

%Density contours for lens models from the Zitrin et al.~and LensPerfect methods
%are overlaid on color images in Figs.~ \ref{fig:SLAdi} and \ref{fig:SLLP}, respectively.

%%%%%%%%%%%%%%%%%%%%%%%%%%%%%%%%%
\begin{figure*}
\epsscale{1.1}
%\plotone{figs/kappa.png}
%\plottwo{figs/kappaAdi.png}{figs/kappaLP.png}
\begin{center}
  \includegraphics[width = .44\textwidth]{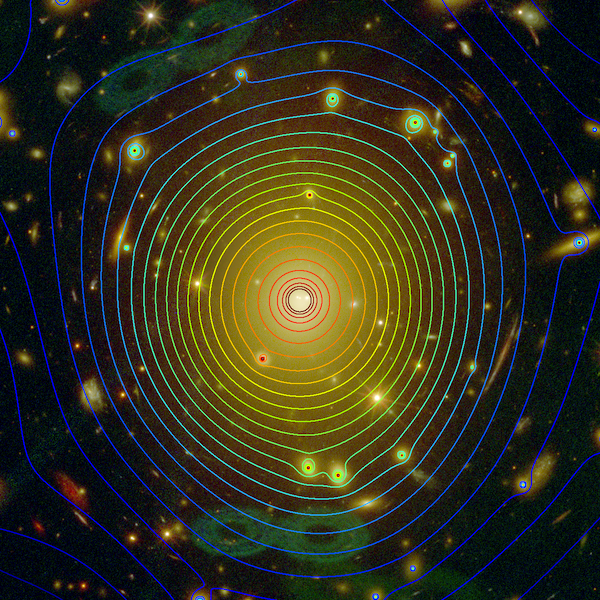}
  \includegraphics[width = .55\textwidth]{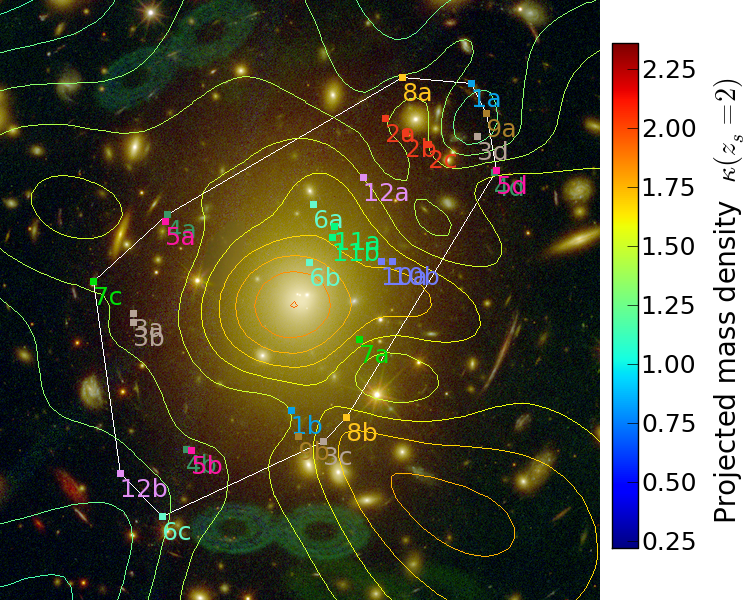}
\end{center}
\caption{\label{fig:SLmap}
Two strong lensing mass models derived for A2261 exhibiting our range of density profile slopes
(plotted in Fig.~\ref{fig:profileM2D}).
The steepest of five models from the \cite{Zitrin09a} method is shown at left
and a shallower model from LensPerfect \citep{Coe08} is shown at right.
The former assumes light approximately traces mass while the latter does not except for a prior on the peak density position.
The LensPerfect model is only well constrained within the white polygon which traces the outermost multiple images.
Plotted are contours of projected mass density in units of the strong lensing critical density ($\kappa = 1$)
for a background object at 
%$z_s = \infty$ ($\sim 2.3 \times 10^{15 }M_{\odot} / \rm{Mpc}^{2}$).
$z_s = 2$ ($\sim 1.9 \times 10^{15 }M_{\odot} / \rm{Mpc}^{2}$).
The contours are logarithmically spaced such that each contour is 10\% denser than the next.
Both ACS+IR color images are to the same scale: $78\arcsec \approx 280$ kpc on a side.
}\end{figure*}
%%%%%%%%%%%%%%%%%%%%%%%%%%%%%%%%%

\section{MMT/Hectospec Spectroscopy}
\label{sec:specz}

\input{figs/tabz}

The Hectospec instrument mounted on the 6.5-meter MMT 
is a multiobject fiber-fed spectrograph with 300 fibers deployable over a 1$^\circ$ diameter field \citep{Fabricant05}.
We targeted probable A2261 cluster members
based on their proximity to the expected cluster red sequence and proximity to the BCG.
In addition, all targets were sufficiently bright with magnitudes $16 < r' < 21$ in SDSS DR7 \citep{SDSS7}.
To observe all targets
we used four pointings 
each with 3$\times$20 minute exposures
within the Hectospec queue schedule. 

After processing and reducing the spectra, we used the Hectospec pipeline \citep{Mink07}
based on the IRAF package RVSAO \citep{KurtzMink98} 
to cross-correlate the spectra with a set of standard Hectospec templates. 
In total, we obtained high quality redshifts for 572 galaxies,
%including 310 within our Subaru analysis region (within $\sim 18'$ of the BCG).
including 308 within our Subaru analysis region (within $17.8'$ of the BCG).
These redshifts are provided in Table \ref{tab:z}.
Scattered light from bright stars in this field limits spectroscopy of this system.
%Table WHICH lists the right ascencion, declination, redshift, and redshift error for the objects acquired with Hectospec. 

Redshift uncertainties are measured for each galaxy individually
and have been empirically quantified globally
% for Hectospec observations 
as follows.
%Empirical uncertainties have been measured as follows.
The SHELS survey \citep{Geller05} carried out with Hectospec
obtained repeat observations of 1468 galaxies,
yielding a mean internal error of
56 km s$^{-1}$ for absorption-line objects and 
21 km s$^{-1}$ for emission-line objects \citep{Fabricant05}.
Comparison of spectroscopic redshifts obtained for 379 galaxies in both SHELS and SDSS DR7
yields $\Delta v = 10 {\rm~km~s^{-1}} \pm 35 {\rm~km~s^{-1}}$.
Note that $\Delta v = 30$ km s$^{-1}$ corresponds to $\Delta z \approx 0.0001$.

%\subsection{A2261 Cluster Members}

Based on the measured redshifts,
we identified cluster members using the caustic technique \citep[e.g.,][]{Diaferio09}.
The technique locates two curves, the caustics, in the cluster redshift diagram,
namely the plane of the line-of-sight velocities of the galaxies 
versus their projected clustercentric distances.
The caustics are related to the escape velocity from the cluster 
and define an area of the redshift diagram where most of the cluster members reside.    
Samples of members identified with the caustic technique 
are at least 95 percent complete
and contaminated by interlopers by 10 percent at most (Serra et al.~2012, in preparation).
This procedure has been used to identify galaxy members of clusters and groups
\citep[e.g.,][]{Rines05,RinesDiaferio06,RinesDiaferio10}
as well as stellar members 
of the Milky Way halo \citep{Brown10} 
and of dwarf spheroidals \citep{Serra10}.

We applied the technique to a sample of 777 spectroscopic redshifts in the field of A2261
(Rines et al.~2012, in preparation).
The technique identifies 209 members within 
%4.6 Mpc/h 
6.6 Mpc $\hh$ of the cluster center. 
We note that the caustic technique also locates the cluster center $\sim 6'$ south of the BCG location 
and at a redshift 0.0017 larger. 
This result indicates that the dynamical structure of A2261 
might be more complex than expected. 
%We will investigate this issue elsewhere. 

%restricted the range of membership selection to 65000-70000 km/s

We used cluster members identified here
to validate galaxy selections based on broad-band photometric colors
used in our strong (\S\ref{sec:SLZitrin}) and weak lensing analyses (\S\ref{sec:background}).

%Hectospec \citep{Fabricant05} on the MMT 6.5-m telescope
%is capable of obtaining spectra for up to 300 objects simultaneously across a diameter of 1 degree.
%Probable cluster galaxies in A2261 were targeted as part of the Hectospec Cluster Survey (HeCS).
%This yielded confident spectroscopic redshifts for 572 galaxies,
%including 308 within $17.8'$ of the BCG (our WL analysis region).
%yielding 638 confident spectroscopic redshifts \citep{Rines10}.

\section{Strong Lens Mass Modeling of the Cluster Core}
\label{sec:SL}

We performed three semi-independent strong lensing (SL) analyses
on the A2261 HST images.
We used the method of \cite{Zitrin09a} to perform the primary strong lensing analysis (\S\ref{sec:SLZitrin})
including the identification of multiple-image systems (\S\ref{sec:SLimages}).
These multiple images are all identified for the first time in this work 
based on our HST imaging and lens modeling.
We verified these identifications using
the Lenstool\footnote{http://www.oamp.fr/cosmology/lenstool/} modeling software \citep{Kneib93thesis,Jullo07}
as well as 
LensPerfect\footnote{\href{http://www.its.caltech.edu/\%7Ecoe/LensPerfect/}{http://www.its.caltech.edu/$\sim$coe/LensPerfect/}}
\citep{Coe08,Coe10},
a ``non-parametric'' method that does not require the assumption that light traces mass (\S\ref{sec:SLother}).
Finally we combined the results from all methods 
yielding an average cluster core mass profile
with uncertainties (\S\ref{sec:SL7}).
By utilizing various modeling methods, 
we captured the true systematic uncertainties
more reliably than generally possible using a single method.

\subsection{Primary Strong Lensing Analysis Method}
\label{sec:SLZitrin}

The \cite{Zitrin09a} method was adapted from that used in \cite{Broadhurst05},
reducing the number of free parameters to six,
and has been used extensively since
\citep{Zitrin09b, Zitrin09a, ZitrinBroadhurst09, Zitrin10A1703, Zitrin11MACS, Zitrin11SDSS, Zitrin11MS1358, Zitrin11A383,
Zitrin11MACS1206, Merten11}.
Basically, the mass model consists of three components:
the cluster galaxies, a dark matter halo, and an external shear
to account for additional ellipticity in the mass distribution in the plane of the sky within or around the core.
Cluster galaxy light is assumed to approximately trace the dark matter;
the latter is modeled as a smoothed version of the former, as described below.

We identified 118 probable cluster galaxies along the ``red sequence''
which is well isolated in F814W-F475W color-magnitude space.
We verified, using additional filters and photometric redshifts, that this selection is robust.
We also compared this selection to Hectospec spectroscopic redshifts
available for 15 galaxies within the HST FOV.
%Thirteen of these are cluster members and two are foreground.
We correctly identified 11 of the 13 cluster members,
missing one near the FOV edge and another near the bright star.
We incorrectly identified one of the two foreground objects ($z = 0.1693$)
as a cluster member as it fell along our red sequence.
These three particular misidentifications have a negligible effect on our mass model
as they all lie at $R > 80''$, 
well outside the strong lensing region where multiple images are formed.
%and beyond the $R = 1'$ outer radius used from our SL model in our joint SL+WL fitting.
Nor do these $\sim 10\%$ rates of incompleteness and contamination
significantly affect our mass profile
as evidenced in part by our other analyses (\S\ref{sec:SL7}).
The cluster members provide a parameterization for the mass model which is not required to be exact
but rather provides a starting point which is molded to fit the data.

Each cluster galaxy is modeled as a power law density profile,
its mass scaling with flux observed in F814W.
%\footnote{This has been shown to work well though F160W may be a better proxy for stellar mass.}
% see notes in /Users/coe/A2744/glens/TIS.py
This mass distribution is then smoothed using a 2D polynomial spline
to provide a model for the dark matter distribution in the cluster halo.
This ``smooth'' mass component is added to the more ``lumpy'' (unsmoothed) galaxy component.
Finally, an external shear is added.
% to account for ellipticity in the mass distribution within or around the core.

In all, there are 6 free parameters: the mass scalings of both the ``smooth'' and galaxy components,
the power law of the galaxy density profiles, the degree of the smoothing polynomial,
and the amplitude and direction of the external shear.
The routine iterates over lens models to find that which best reproduces 
the observed positions of the strongly lensed images
{\em in the image plane}, rather than in the source plane, 
which can bias solutions toward flatter profiles and higher magnifications.
Complete details may be found in \cite{Zitrin09a}.

In this paper, we introduce an alternative Gaussian convolution kernel
%Gaussian instead of polynomial spline --
to produce the ``smooth'' mass component.
In this case, the free parameter is the Gaussian width
instead of the polynomial spline degree.  %interpolation

\subsection{Multiple Images of Strongly Lensed Galaxies}
\label{sec:SLimages}

Using this method, we identified 30 multiple images of 12 background galaxies strongly lensed by A2261
(see Figures \ref{fig:arcs} and \ref{fig:arcstamps} and Table \ref{tab:arcs}).
These are all identified for the first time in this work.
%(though the bright arc was easily identifiable previously).

We used an iterative process to identify images and add them to the model,
beginning with those which are most confident.
Our most confident multiple image system is the ``claw'' or U-shaped object, system 1.
%with a photometric redshift $z \sim 4.4$.
The distinctive morphology is apparent in both images,
including a color gradient best viewed in the IR color images with the BCG subtracted (see Fig.~\ref{fig:arcstamps}).
Image 1a yields a photo-z $z \sim 4.4$.
The IR flux of image 1b appears to be biased a bit high by contaminating light from the BCG (Fig.~\ref{fig:sedcompare}),
such that the best fit SED is an early type galaxy at $z \sim 0.5$.
The irregular morphology is not consistent with an early type galaxy.

Based on this system, we obtained the initial mass model,
enabling us to predict the lensed positions of counterimages of other galaxies
by delensing them to their putative true source positions and then relensing them with our model.
Candidate counterimages were identified as being near the observed position,
with the predicted lensed morphology and orientation
as well as consistent colors and photometric redshifts.
Observed photometry and SED fits are shown in Fig.~\ref{fig:sedcompare}.
Our multiple images generally have consistent observed SEDs
(allowing for variations in magnification) and thus photo-z's.
However some images yield unreliable photo-z's if they are faint and/or 
their light is contaminated by a bright nearby cluster member.
To date, no spectroscopic redshifts are available for these galaxies strongly lensed by A2261.

Our 6-parameter model is fully constrained by the positions of our 30 multiple images.
%enabling us to accurately determine the mass distribution and inner profile,
%contingent on the assumption that this parametrization yields an accurate description of the mass distribution.
%Five final mass models were produced, spanning a range of parameters and use various smoothing procedures.  
%They bracket the range of acceptable solutions given this parameterization.
We produced four mass models spanning the range of profile slopes allowed by the data
as dictated by the density power law
%($\rho \propto r^{-\gamma}$)
of the cluster galaxies ($1.1 \leq q \leq 1.2$, where the surface density $\Sigma \propto R^{-q}$)
and the polynomial smoothing degree ($4 \leq S \leq 8$).
We also generated a fifth mass model using a Gaussian convolution kernel
with an optimized width of $9.1''$ and $q = 1.3$.
This model is shown in Fig.~\ref{fig:SLmap}.

We note that our lens models do not predict counterimages for the large prominent arc
marked with an ``X'' in Fig.~\ref{fig:arcs}.
%along the right side of Fig.~\ref{fig:arcs}.
Instead they predict a single highly-distorted image, as observed.
%It appears to be a single galaxy with a highly distorted image.
We measure its photometric redshift to be $z = 1.19^{+0.05}_{-0.02}$ (95\% C.L.).
If it were at a slightly higher redshift $z \sim 1.5$,
some of our models would predict a radial arc counterimage
on the opposite side of the BCG core.
We detect no such image.
This search is aided by the fact that the arc is significantly detected in F390W
where most of the other arcs drop out and the BCG light is significantly reduced.
% ACSIR:  1.189 +0.049 -0.022
% 2975 1.1890 1.167  1.238  7.750   0.984972 1.185 7.750      3.773  22.864

We initially identified a possible counterimage to this large arc
with similar colors and photo-z
at RA, Dec (J2000) = 17:22:29.4, +32:07:29 
(near the bottom left corner of Fig.~\ref{fig:arcs}).
However, the lens models required to reproduce this counterimage
were significantly stronger (higher mass) than our final models described above
and thus inconsistent with all of our multiple image systems.
They also predicted an additional multiple image to the North which is not observed
(in the vicinity of 17:22:27.8, +32:08:16).

\subsection{Complementary Strong Lensing Analyses}
\label{sec:SLother}

We performed semi-independent lens modeling analyses 
using Lenstool \citep{Kneib93thesis,Jullo07}
and LensPerfect \citep{Coe08,Coe10}.
In the course of these analyses,
we verified the multiple image systems
and estimated their redshifts independently.

Our Lenstool model consisted of an NFW halo \citep{NFW96}
%(Appendix \ref{sec:appendix})
and truncated PIEMD (pseudo-isothermal elliptical mass distribution) halos \citep{KassiolaKovner93}
%SIS (singular isothermal sphere) halos 
%(PIEMDs, or pseudo-isothermal elliptical mass distributions)
for the 69 brightest cluster members, which were again identified photometrically
but independently from the analysis in \S\ref{sec:SLZitrin}.

For the LensPerfect analysis,
we assumed a prior that the mass is densest near the center of the BCG
and roughly decreases outward radially.  
Otherwise, it includes no assumptions about light tracing mass.
Other priors include overall smoothness and rough azimuthal symmetry.
For details, see \cite{Coe08,Coe10}.
%One resulting mass model solution is shown in Fig.~\ref{fig:SLLP}.
The best solution found, according to these criteria, is shown in Fig.~\ref{fig:SLmap}.
It perfectly reproduces the observed positions of all multiple images (to the accuracy with which they are input).
%A range of such solutions are possible, and this was the ``most physical'' 
%found according to a set of non-restrictive criteria described in detail in \citep{Coe08}.
%In addition to the above priors, minimize excessive variations.
We note the solution is only well constrained within the white polygon which bounds the multiple images.

%%%%%%%%%%%%%%%%%%%%%%%%%%%%%%%%%
\begin{figure}
\epsscale{1.1}
%\plotone{figs/massprofileprojected7SLzoom.png}
\plotone{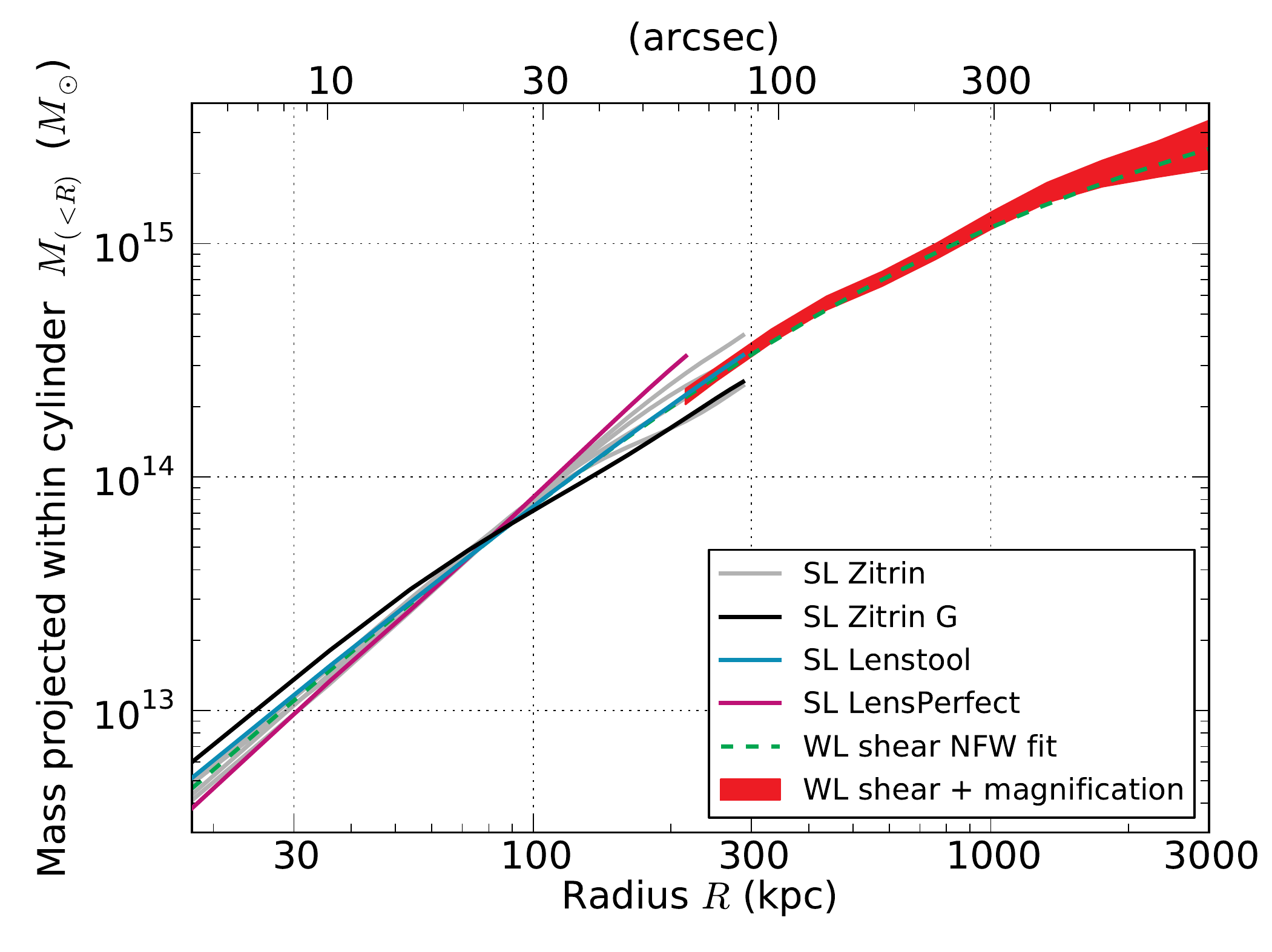}
\caption{\label{fig:profileM2D}
%XXX Plot will be redone. 
Projected 2D mass profiles from strong and weak lensing.
We present seven SL models and two WL models.
Five of the SL models are based on the Zitrin method (\S\ref{sec:SLZitrin}).
One of these (``Zitrin G'') uses a different (Gaussian) kernel 
to smooth the light distribution for use as the halo mass model.
The LensPerfect model is steepest in integrated $M(<R)$,
which translates to the shallowest density profile $\kappa(R)$,
while the Lenstool model is about average (\S\ref{sec:SLother}).
}\end{figure}
%%%%%%%%%%%%%%%%%%%%%%%%%%%%%%%%%

%%%%%%%%%%%%%%%%%%%%%%%%%%%%%%%%%
\begin{figure}
\epsscale{1.1}
\plotone{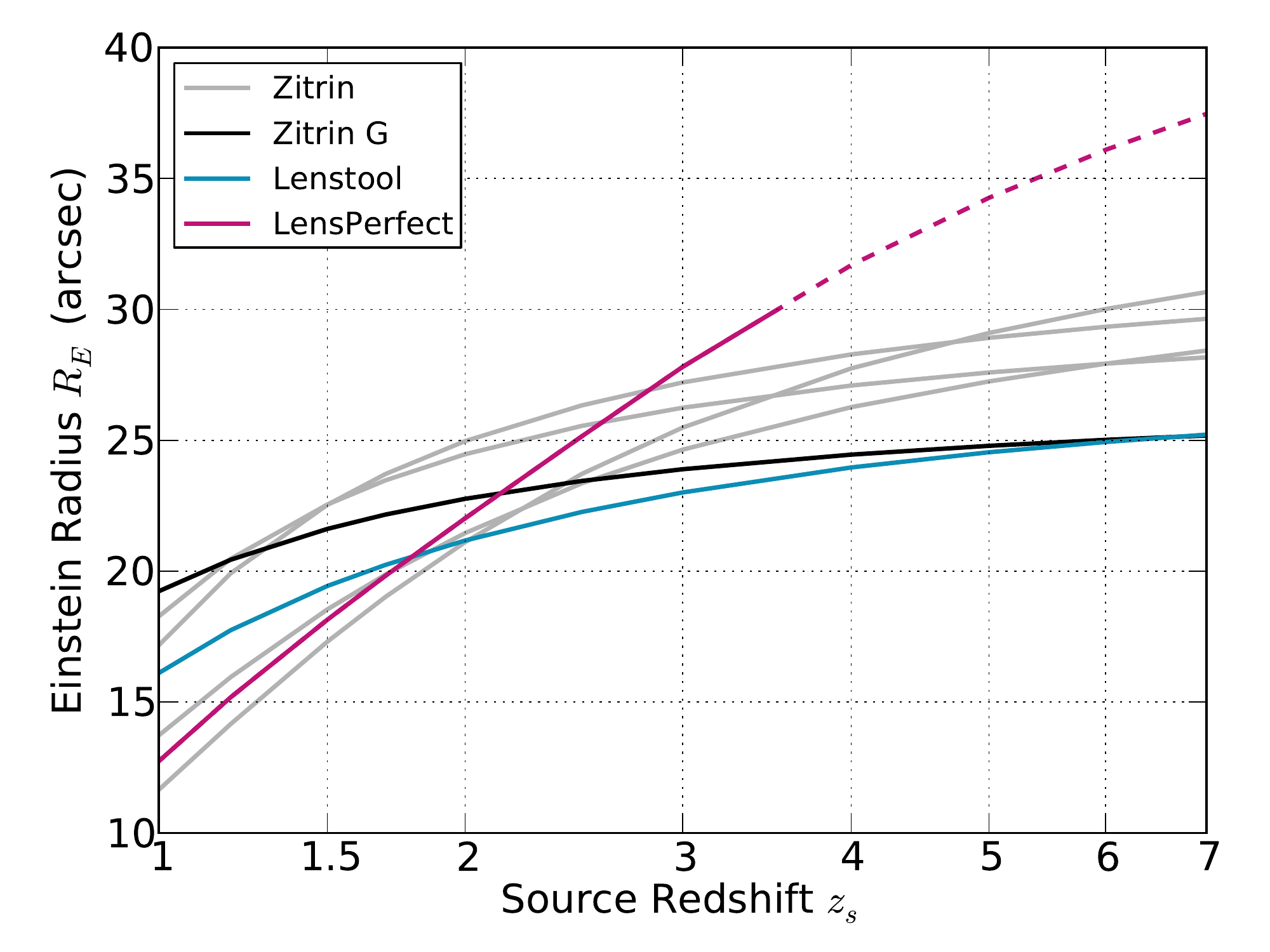}
\caption{\label{fig:REzs}
Einstein radius $R_E$ as a function of source redshift $z_s$
measured by our various lens models
(with mass profiles plotted in Fig.~\ref{fig:profileM2D}).
Our results are much lower than the rough estimate of
$R_E \sim 40\arcsec$ $(z_s = 1.5)$
from ground-based imaging
quoted by \cite{Umetsu09}.
Instead, our models range between
$R_E = 20\arcsec \pm 2\arcsec$ $(z_s = 1.5)$ and $23\arcsec \pm 2\arcsec$ $(z_s = 2)$.
% 17.5 - 22
% 21 - 25
Some level of agreement is guaranteed by the fact that 
all models used the same
input multiple images and photometric redshift information,
though the redshifts are allowed to vary somewhat 
and are optimized independently by each method.
The LensPerfect mass profile is shallow, hovering near the lensing critical density,
allowing unconstrained perturbations beyond $R \gtrsim 30''$
(including some area outside the multiple image constraints)
to significantly influence $R_E$ at these radii.
}\end{figure}
%%%%%%%%%%%%%%%%%%%%%%%%%%%%%%%%%

\subsection{Results from the Ensemble of Models}
\label{sec:SL7}

Integrated projected mass profiles from our 7 SL mass models are presented in Fig.~\ref{fig:profileM2D}.
We adopt the average and scatter of these models as our final SL constraints.
Some level of agreement is guaranteed by the fact that 
all models used the same input multiple images and photometric redshift information.
These image identifications were verified independently in each analysis.
The redshifts were allowed to vary somewhat and were optimized independently by each model.

The models converge most tightly on the projected mass contained within $\sim 20\arcsec$,
roughly as expected given the Einstein radii of the systems.
In Fig.~\ref{fig:REzs}, we plot the Einstein radius $R_E$ as a function of background source redshift $z_s$ for all 7 SL models.
We calculated these $R_E(z_s)$ as those circular radii centered on the highest mass peak 
(coincident with the BCG)
which enclose an average projected density equal to the critical strong lensing density.

Our results, which range from
$R_E = 20\arcsec \pm 2\arcsec$ $(z_s = 1.5)$ and $23\arcsec \pm 2\arcsec$ $(z_s = 2)$,
are significantly lower than those roughly estimated from ground-based imaging
as quoted by \cite{Umetsu09}: $R_E = 40\arcsec \pm 4\arcsec$ $(z_s = 1.5)$.
This previous estimate was based on 
$R_E \sim 30''$ for $z_s \lesssim 1$
assuming that the bright, prominent arc 
(marked with an X in Fig.~\ref{fig:SLmap})
lies near the Einstein radius.
In this work, we find this arc is not in fact located at $R_E$.
Instead, it lies at $R \approx 27''$, greater than the $R_E = 17'' \pm 3''$ ($z_s = 1.2$)
determined robustly by our 12 other multiple image systems.
Furthermore, we find it has a photo-z $\sim 1.2$, greater than that assumed in \cite{Umetsu09}.
Both of these factors contributed to the higher concentration measured in that work.

\section{Weak Lensing Mass Modeling}
\label{sec:WL}

To probe the mass distribution of A2261 at larger radii, 
we turn to weak lensing analyses of wider ground-based images
obtained with Subaru Suprime-Cam,
as previously studied in detail by \cite{Umetsu09} and \cite{Okabe10}.
Here we present new, more robust analyses incorporating 
additional observations, improved image reductions, and new analytical techniques.
The additional observations include KPNO Mayall 4-m imaging
(\S\ref{sec:widefield})
and spectroscopy from MMT/Hectospec
(\S\ref{sec:specz}).

%\subsection{Observations: Wide Field Imaging}
%\subsection{Multiband Subaru Wide-Field Imaging}
%\subsection{Subaru Wide-Field Imaging}
%\subsection{Wide-Field Imaging and Spectroscopy}
\subsection{Subaru and KPNO Wide-Field Imaging}
\label{sec:widefield}

%obtained with the $34\arcmin \times 27\arcmin$ FOV Suprime-Cam \citep{Miyazaki02} on the 8.2-m Subaru telescope,
%NOAO archival 

Our weak lensing analysis is based on archival
\BVR\ imaging
obtained with the
$34\arcmin \times 27\arcmin$ FOV 
Suprime-Cam \citep{Miyazaki02}
on the Subaru 8.2-m telescope,
%$34\arcmin \times 27\arcmin$ FOV Subaru Suprime-Cam \citep{Miyazaki02} \BVR\ imaging
%and NOAO archival KPNO Mayall 4-m MOSAIC1 \iz\ imaging (program 2008A-0356, P.I. Mandelbaum).
and NOAO archival MOSAIC1 \iz\ imaging obtained with the Mayall 4-m telescope at KPNO
(program 2008A-0356, P.I. Mandelbaum).
The integration times are 20, 30, 45, 60, and 90 minutes, respectively, for the five filters \BVRiz.
%The $34\arcmin \times 27\arcmin$ FOV Suprime-Cam \citep{Miyazaki02} on the 8.2-m Subaru telescope,
%
The Subaru \RC-band images used for
galaxy shape analyses
have a seeing FWHM $\approx 0.65\arcsec$
and were obtained at two different orientations rotated by 90 degrees
(see \S\ref{sec:shape}).

%\subsection{Previous Weak Lensing Analyses}
%\subsection{Weak Lensing Mass Profile: Previous Results}
\subsection{Previous Analyses}
\label{sec:WLprev}

%To probe the mass distribution of A2261 at larger radii, 
%we turn to weak lensing analyses of wider ground-based images from the Subaru SuprimeCam.
\cite{Umetsu09} and \cite{Okabe10} 
both measured weak lensing distortions in the Subaru \RC-band imaging described above.
Background galaxies were selected based on \VR\ color-magnitude cuts
determined in part so as to minimize contamination of unlensed cluster galaxies
as described in \cite{Medezinski07} and \cite{UmetsuBroadhurst08}.

Both analyses identified 
a massive background structure at $z \sim 0.5$
which may affect the lensing signal.
\cite{Umetsu09} found that out of four clusters analyzed,
A2261 was the most sensitive to the exact profile fitting method used.
% at radii of $4\arcmin$ and $10\arcmin$ from the BCG.
% about $7\arcmin$ south of the BCG.

The first method, used in both papers (and commonly elsewhere),
fits the observed shears (binned radially) directly to those expected from NFW profiles.
The second method 
%is more robust in the face of additional line-of-sight structures.
attempts to correct for the mass-sheet degeneracy based on the observed shears alone.
%The discretized density profile is iteratively refined with the outer ``mass sheet'' density left as a free parameter,
With the outer ``mass sheet'' density left as a free parameter $\kappa_b$,
the discretized density profile is iteratively refined
toward consistency with the observed reduced tangential shears $g_+ = \gamma_+ / (1 - \kappa)$.
This method \citep{UmetsuBroadhurst08,Umetsu09} 
is a non-linear extension of earlier ``aperture densitometry'' techniques 
developed by \cite{Fahlman94} and \cite{Clowe00}.

\cite{Umetsu09} found that NFW fits to the profile derived from the latter method
yield a marginally higher mass concentration 
$c_{vir} = 10.2^{+7.1}_{-3.5}$ than the $c_{vir} = 6.4^{+1.9}_{-1.4}$ obtained using the former method
(Table \ref{tab:Mc} and Fig.~\ref{fig:massprofileprojected}).
This higher value was supported by their rough estimate of the strong lensing
Einstein radius $R_E \sim 40\arcsec$ ($z_s \sim 1.5$) based on the prominent arc (see \S\ref{sec:SL7}).
%A joint fit to the WL + $R_E$ 
A joint fit to the WL and $R_E = (40 \pm 4) \arcsec$ 
%% KU@@
%($R_E\sim 40\arcsec$) 
yielded $c_{vir} = 11.1^{+2.2}_{-1.9}$,
which was quoted by \cite{Oguri09} as an example of a cluster with higher than expected concentration.

%In \S\ref{sec:SL} we revisited the SL measurement, finding a lower value of $R_E = (20 \pm 2) \arcsec$ for $z_s = 1.5$.
%Here we also reanalyze the \cite{Umetsu09} WL data
%to better understand the origin of the $c_{vir} \sim 10$ WL result.

We revisit the \cite{Umetsu09} $c_{vir} \sim 10$ WL result as follows.
In that analysis, the innermost mass density profile point $\bar{\kappa}(<1\arcmin)$ 
(corresponding to the mean density interior to the inner radial boundary of weak lensing measurements) 
was not included in the fitting.  
Here we included that data point and found the best fit value and uncertainties both decreased
from $c_{vir} = 10.2^{+7.1}_{-3.5}$ to $c_{vir} = 5.8^{+1.8}_{-1.4}$.
We note the significant improvements in precision gained by increasing the radial range of constraints.

\cite{Okabe10} found a lower virial mass 
$M_{vir} = 1.36^{+0.28}_{-0.24} \times 10^{15} M_\odot$
and
$c_{vir} = 6.04^{+1.71}_{-1.31}$.
This profile underestimates mass in our strong lensing region by $\sim 20\%$ (\S\ref{sec:lensing}).
We compared our shear catalog directly with that used in \cite{Okabe10} and provided to us.
We find similar WL signal for $R > 3'$
but recover stronger signal interior to this radius.
This difference is most likely explained by
improved background selection in our catalog 
with lower contamination due to cluster galaxies (\S \ref{sec:background}).

\subsection{Current Analysis}
\label{sec:WLcurrent}

%We utilize one additional Subaru band (\BJ) 
In addition to the Subaru \VJ\ and \RC\ images,
we also utilize the \BJ-band image,
which improves our selection of background galaxies (\S\ref{sec:background})
with respect to the previous analyses.
Our Subaru image reduction procedure \citep{Nonino09}
is somewhat improved compared to that used in \cite{Umetsu09}
in terms of distortion corrections and image co-addition (here PSF-weighted).
%**you need to add some details about Mario's reduction here. You can see from his 2009 paper**
After trimming the shallower edges, 
the final co-added images roughly cover a circular area with a $17.8'$ radius
($\sim$ 1,000 square arcmin)
which we use for our analysis.
As in the previous analyses
we measured galaxy shapes in the Subaru \RC-band images,
though our procedure is slightly different
(\S\ref{sec:shape}).

%(IDs K1019 and K1020, respectively).
The KPNO \ip\ and \zp\ images were reduced using calibration frames, 
including fringe and pupil maps, obtained from the archive,
and then stacked in a manner similar to the Subaru images.
Zeropoints were calibrated based on comparisons 
with point source photometry from SDSS DR7 \citep{SDSS7}.

Five-band \BVRiz\ photometry was measured using SExtractor
in PSF-matched images created by ColorPro \citep{Coe06}.
Subaru zeropoints were calibrated based on comparisons to HST and KPNO photometry,
%(the latter calibrated to SDSS)
then recalibrated based on SED fits to photometry of galaxies with measured spectroscopic redshifts
primarily from Hectospec and supplemented by SDSS DR7 (\S\ref{sec:specz}).

\subsection{Shape Measurement}
\label{sec:shape}

We produced two separate co-added \RC-band images for the shape analyses
based on the imaging obtained at two different orientations separated by 90 degrees.
%We computed weighted averages of the galaxy shapes (reduced shears) measured at each orientation.
Galaxy shapes (reduced shears) were measured at each orientation
and their weighted averages computed.
In both \cite{Umetsu09} and \cite{Okabe10},
shapes were instead measured in co-added images which combined both orientations.
We find this does not have a significant effect on the derived mass profile.
When we use the same method to
analyze shears measured in \cite{Umetsu09} and in this work,
we find consistent results (Table \ref{tab:Mc}, rows 5 and 6).

%For that analysis, galaxies were detected and their PSF-corrected shapes were measured 
%using IMCAT and a KSB-based approach \citep{KSB,Erben01}.
%%%%
%KU@@ (2011/11/21)
For accurate shape measurements of faint background galaxies, 
%we used the IMCAT package developed by N. Kaiser,
%to perform object detection and shape measurements,
%we used IMCAT following the formalism outlined in \cite{KSB}.
we used the IMCAT software \citep{KSB}, 
following the formalism outlined in that paper.
%Kaiser, Squires \& Broadhurst (1995, hereafter, KSB). 
Full details of our weak lensing analysis pipeline are provided in \cite{Umetsu10}.
We have tested our shape measurement and object selection pipeline using
simulated Subaru Suprime-Cam images 
(M. Oguri 2010, private communication; \citealt{Massey07}).  
We recover input WL signals with good precision:
typically a shear calibration bias $|m| \lesssim 5\%$
%% 2011/11/25 @@KU
(where this bias shows a modest dependence of calibration accuracy on seeing conditions),
and a residual shear offset $c\sim 10^{-3}$,
which is about one order of magnitude smaller than the typical distortion signal 
(reduced shear $|g|\sim 10^{-2}$) in cluster outskirts.
This level of performance is comparable to other similarly well-tested methods \citep{Heymans06},
and has been improved in comparison with our previous pipeline used in \cite{Umetsu09}
which achieved 
$5\% \lesssim |m| \lesssim 10\%$.
%$-10\% \lesssim m \lesssim -5\%$.

\subsection{Background Galaxy Selection}
\label{sec:background}

Robust selection of background galaxies is crucial in weak lensing analyses
to minimize contamination by unlensed cluster and/or foreground galaxies
which would dilute the lensing signal by a fraction equal to the level of contamination
\citep{Broadhurst05b,Medezinski07}.
If not accounted for properly,
contamination can be especially significant at small clustercentric radii
where cluster galaxies are relatively dense.
Previous analyses have demonstrated that 
color-color selection using three Subaru broadband filters
delivers robust discrimination between cluster, foreground and background galaxies
\citep{Medezinski10,Medezinski11,Umetsu10,Umetsu11,Umetsu11b}.

%%%%%%%%%%%%%%%%%%%%%%%%%%%%%%%%%
\begin{figure}
\epsscale{1.1}
%\plotone{figs/A2261_BVR_samples_111114.pdf}
%\plotone{figs/A2261_BVR_samples_111205.png}
\plotone{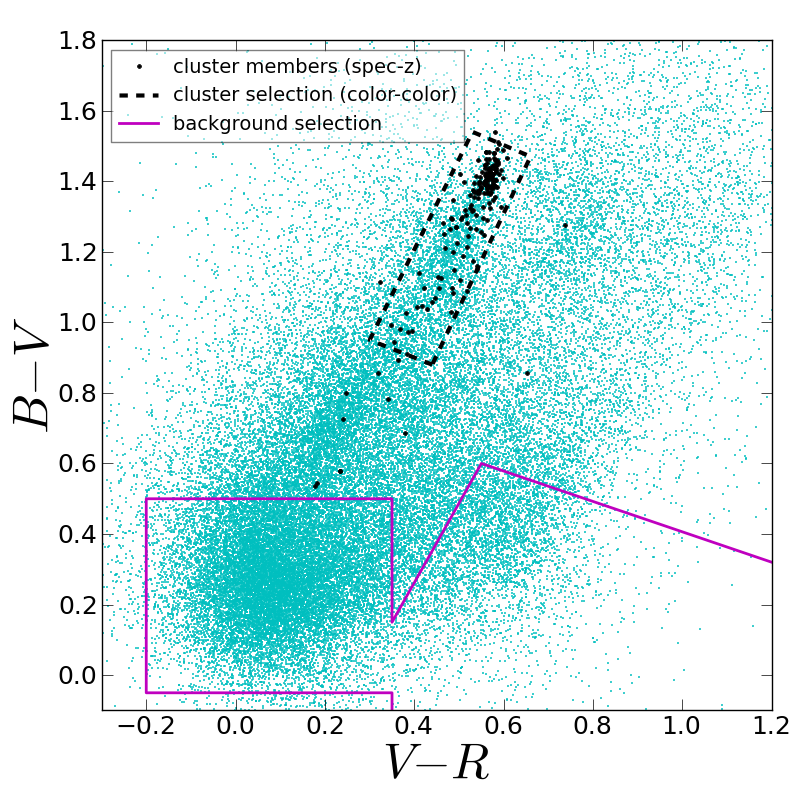}
\caption{\label{fig:CC}
Background galaxies are selected for weak lensing analysis
(lower outlined region)
based on Subaru \BVR\ color-color-magnitude selection.
At small radius, the cluster overdensity is identified as the dashed region.
The background samples are well isolated from this region
and satisfy other criteria
as discussed in \S\ref{sec:background}.
This background selection successfully excludes all 189 cluster members (black) within the Subaru FOV
as identified based on our velocity caustic analysis of Hectospec spectroscopic redshifts (\S\ref{sec:specz}).
Cyan points are $\RCm < 26$ galaxies, where stars have been identified and excluded.
%
%(Colors in this plot will be updated to avoid ``red'' and ``blue''.)
}\end{figure}
%%%%%%%%%%%%%%%%%%%%%%%%%%%%%%%%%

%Here we began by detecting objects within our circular ($17.8'$ radius) field.
Here we began by detecting objects within $17.8'$ ($\sim 3.8$ Mpc) of the BCG
(the area deeply imaged by Subaru).
We pruned stars from this sample
based on \RC-band magnitude, peak flux, FWHM, and SExtractor ``stellarity''.

We then derived \BJ\VJ\RC\ color-color-magnitude cuts (Figure \ref{fig:CC})
as described in \cite{Medezinski07,Medezinski10,Medezinski11}.
%
%In this color-color space,
We calculated number count density and average clustercentric radius
both as a function of position in this color-color space.
Cluster galaxies are identified as a peak in the former and minimum in the latter.
We determined the region occupied by these galaxies
and later found it to coincide well with colors of cluster members 
as determined based on a velocity caustic analysis of Hectospec spectroscopic redshifts (\S\ref{sec:specz}).

%Background galaxies are identified as other number count peaks.
We then defined regions in this color-color space well separated from the cluster galaxies
for use as our background galaxy selection.
The border placement is optimized
to maximize total number counts while minimizing contamination from cluster members.
The latter can be detected as dilution of the average shear signal
and/or
a rise in number counts toward the cluster center.
% positional clustering near the BCG.
% decrease in number counts as a function of clustercentric radius.
We also imposed magnitude cuts $22 < \RCm < 26$
to further avoid contamination at the bright end and incompleteness at the faint end,
while maximizing the number of faint galaxies which contribute to the lensing signal.

Our final cuts (Fig.~\ref{fig:CC}) yielded 12,762 background galaxies 
(12.8 $/$ square arcmin) for WL analysis.
We verified that the WL shear signal increases toward the center
and that the B-mode (curl component) is consistent with zero.
We later verified that these cuts successfully reject all 189 galaxies
identified spectroscopically as cluster members within the Subaru FOV (\S\ref{sec:specz}).
%based on a velocity caustic analysis of Hectospec spectroscopic redshifts (\S\ref{sec:specz}).
%identified within the Subaru FOV
%based on a velocity caustic analysis of Hectospec spectroscopic redshifts (\S\ref{sec:specz}).

To estimate the mean effective redshift of this background population,
we applied this same color-color-magnitude cut to 
galaxies with robust photometry and photometric redshifts measured in the COSMOS field
\citep{Capak07,Ilbert09}.
We compute the average lensing efficiency $\beta = D_{LS} / D_{S}$ for this sample
given our lens redshift $z_{L} = 0.225$
%for the $YY\%$ of galaxies at $z < 0.224$ which would be unlensed by A2261.
and find an effective $z_{S} = 0.99\pm0.10$ for our background WL sample.
For each lensed galaxy, the factor $\beta$ is a function of angular diameter distances
from lens to source $D_{LS} = D_A(z_L,z_S)$
and observer to source $D_S = D_A(0,z_S)$.
We later marginalized over this uncertainty when fitting mass profiles to our WL data.
%computing \Mvir\ and \cvir.

%%%%%%%%%%%%%%%%%%%%%%%%%%%%%%%%%
\begin{figure}
\epsscale{1.1}
\plotone{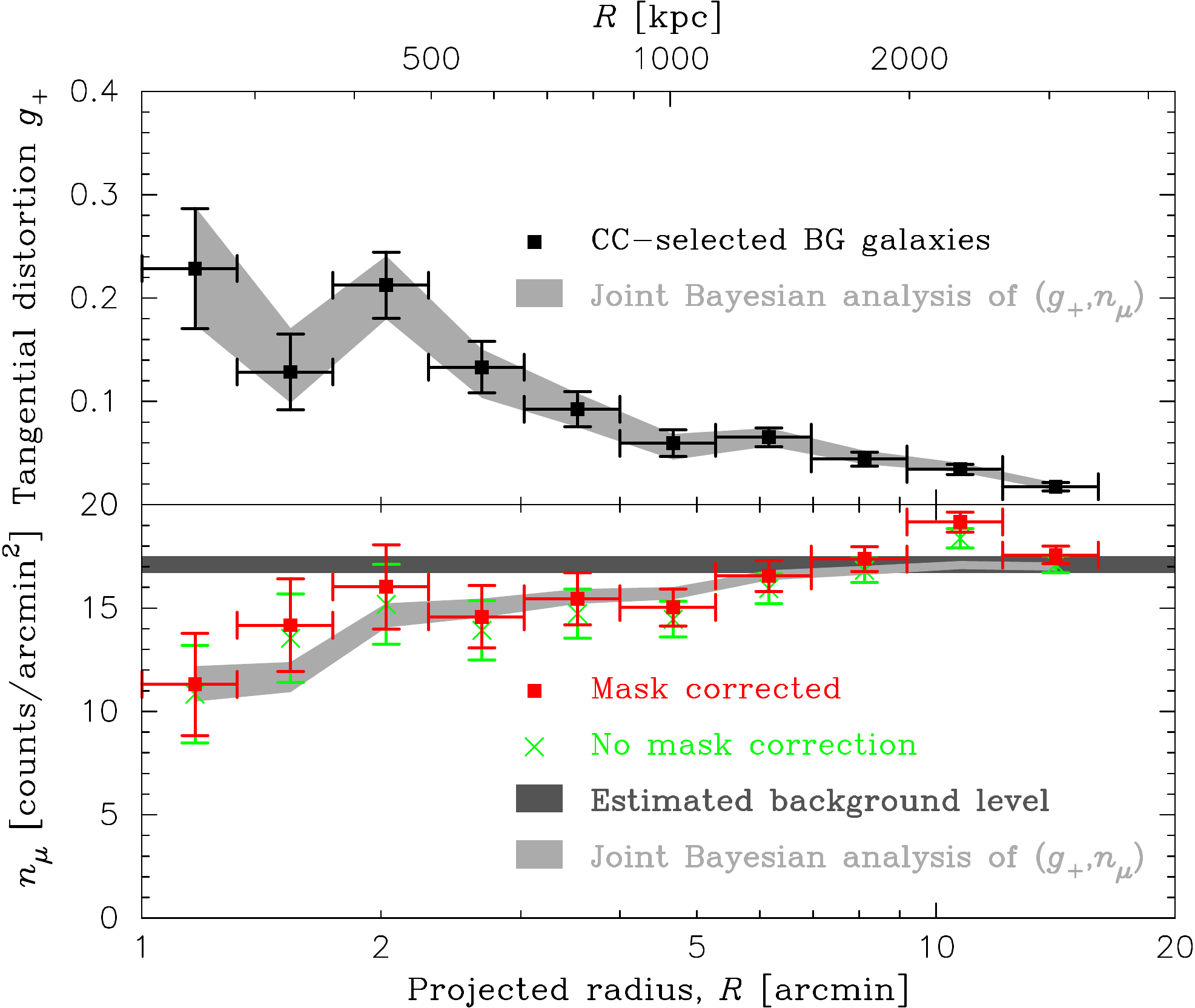}
\caption{\label{fig:WLprofile}
Weak lensing tangential reduced shear (\emph{top})
and magnification (number count depletion) profiles (\emph{bottom})
measured in Subaru images of A2261
(\S\ref{sec:WLprofile}).
Also plotted is a joint Bayesian fit to both.
This is our primary WL model.
Bright objects are masked out to refine the estimates of area and thus number count densities.
}\end{figure}
%%%%%%%%%%%%%%%%%%%%%%%%%%%%%%%%%

%\subsection{Weak Lensing Mass Profile}
\subsection{Revised Weak Lensing Mass Profile}
%\subsection{Revised Lower Concentration}
\label{sec:WLprofile}

In addition to our revised selection of background galaxies,
we also used a slightly different method to estimate the ``mass-sheet'', or background density $\kappa_b$.
Here we performed iterative NFW fitting allowing $\kappa_b$ to be a free parameter \citep{Umetsu10}.

%We then perform a second analysis method which incorporates weak lensing magnification data
%(depleted number counts of background galaxies).
%For details on this method, see \cite{Umetsu11}.
%Measurements of number count depletion generally break the mass sheet degeneracy more robustly \citep[e.g.,][]{Broadhurst95}
%than the aperture densitometry technique described above.
%A joint Bayesian fit to both the observed shears and magnification is shown in Fig.~\ref{fig:WLprofile}.
%% KU@@
We then performed a second analysis method which incorporates independent
weak-lensing magnification data 
(depleted number counts of faint background galaxies) in a Bayesian approach.
%({\bf KU: Are we going to show background galaxy statistics,
%say the total number of galaxies, the number density, redshift and
%lensing depth $\beta$ information, etc.?}).
%% KU@@
For details on this method, see \cite{Umetsu11,Umetsu11b}.
Measurements of number count depletion generally break the mass sheet
degeneracy more robustly \citep[e.g.,][]{Broadhurst95,Umetsu11} 
than the aperture densitometry technique described above.

We find a consistent mass profile solution 
based on a joint Bayesian fit to both the observed shears and magnification
as shown in Fig.~\ref{fig:WLprofile}.
The total signal-to-noise ratio (S/N) in our tangential distortion profile is ${\rm S/N}\approx 17$ 
(defined as in equation 9 of \citealt{Okabe10}), 
whereas ${\rm S/N}\approx 20$ in the joint mass profile from
combined tangential reduced shear and magnification measurements
%% 2011/11/25 KU@@
(equation 38 of \citealt{UmetsuBroadhurst08}).
Thus, in addition to breaking the mass sheet degeneracy,
the magnification measurements also increased the overall significance by $\sim 20\%$
%% 2011/11/25 KU@@
(cf.~Table 5 of \citealt{Umetsu11b}; \citealt{RozoSchmidt10}).
%Rozo & Shmidt 2011, arXiv:1009.573).
%gain $\sim 20\%$ in WL signal.
%Thus the improvement here from adding the magnification measurements is
%$\approx 20\%$, compared with the weak lensing shear signal.
%This high significance is comparable to other strong lensing clusters analyzed similarly in \cite{Umetsu11}.
%we find A2261 would be second highest. }
%{\bf Comparing this to the five clusters analyzed in \cite{Umetsu11},
%we find A2261 would be second highest. }

Our \BVR\ color-color selection does not allow us to effectively discriminate
between ``blue'' and ``red'' background samples
with properties similar to those derived from \BRz\ color-color selection \citep{Medezinski10,Medezinski11}.
Galaxies in the ``blue'' samples identified in these works
have steep number count slopes, roughly canceling out any number count depletion.
Stronger magnification signals are measured in ``red'' \BRz\ samples
with relatively flatter number counts.

To investigate the effect this may have on our analysis,
we explored the \BRz\ colors of
the subset of our galaxies detected in shallower \zp-band KPNO imaging.
We found that the majority of our background sample corresponds to a ``red'' selection in \BRz, as desired.
We repeated our magnification analysis on this red subset and found no significant changes in our results
except for somewhat larger uncertainties due to the lower number of galaxies.

%%%%%%%%%%%%%%%%%%%%%%%%%%%%%%%%%
\begin{figure}
\epsscale{1.17}
%\plotone{figs/kappaprofile.pdf}
%\plotone{figs/massprofileprojected7.png}
%\plotone{figs/kappaprofile9.pdf}
%\plotone{figs/kappaprofileb.pdf}
\plotone{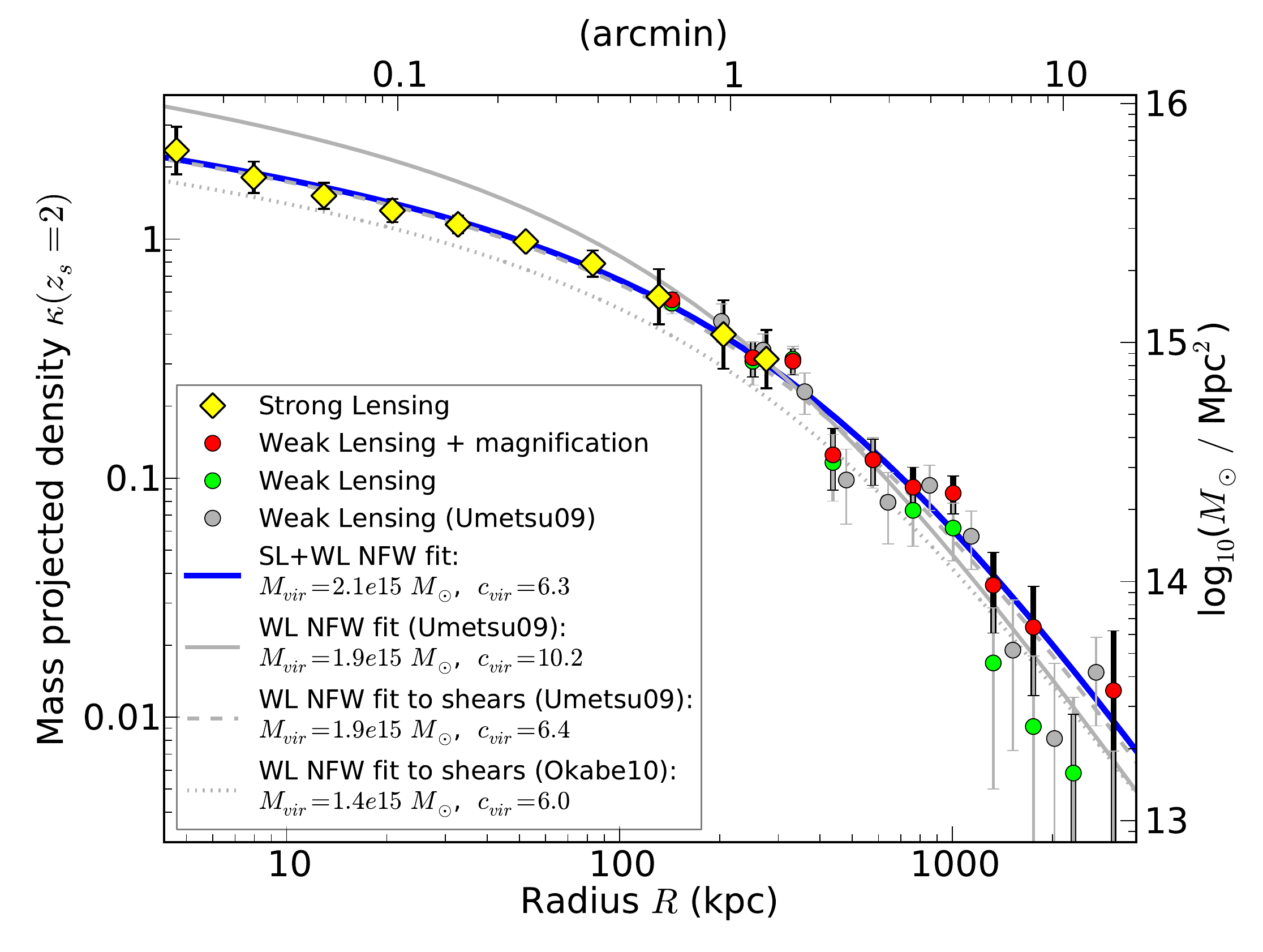}
\caption{\label{fig:massprofileprojected}
Projected mass density profile derived from strong and weak lensing analyses
fit to NFW profiles from published analyses (gray) and this work (colors).
The SL results are the average and scatter of our seven mass models (\S\ref{sec:SL7}).
}\end{figure}
%%%%%%%%%%%%%%%%%%%%%%%%%%%%%%%%%

%%%%%%%%%%%%%%%%%%%%%%%%%%%%%%%%%
\begin{figure}
\epsscale{1.2}
%\plotone{figs/kappaprofile.pdf}
%\plotone{figs/Mc7.png}
%\plottwo{figs/MckappaHD.png}{figs/McMCMC.png}
%\plotone{figs/McMLE5_121.png}
%\plotone{figs/Mckappa3_121.png}
%\plotone{figs/Mckappa3_121.pdf}
%\plotone{figs/Mc33.pdf}
\plotone{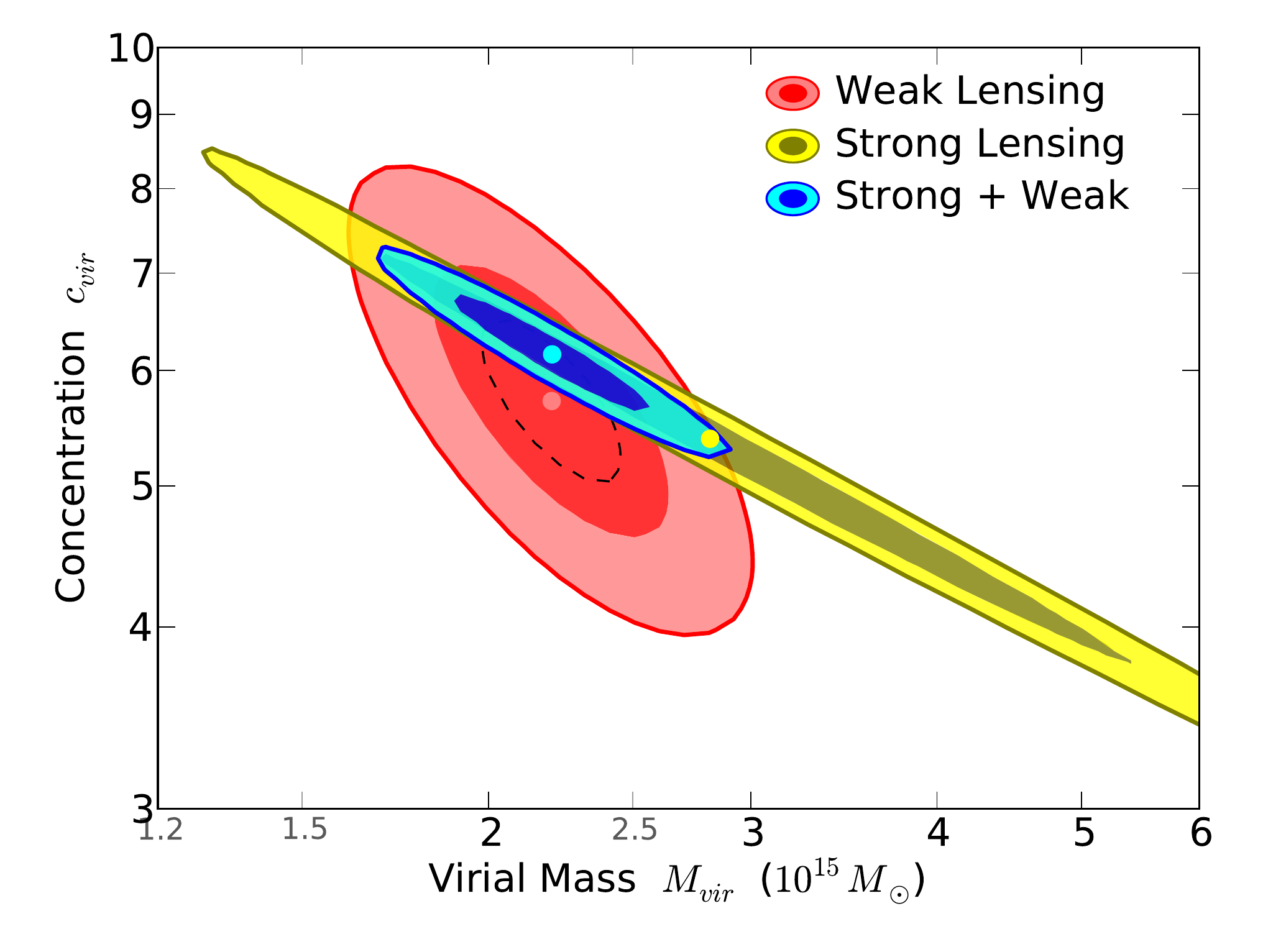}
\caption{\label{fig:M-c}
Constraints on the virial mass and concentration of A2261 from lensing analyses.
Contours are 1-$\sigma$ and 2-$\sigma$ ($\Delta \chi^2 = 2.30$ and 6.17, 
or $\approx 68\%$ and 95\% confidence assuming Gaussian ellipsoidal uncertainties).
The dashed black line is 1-$\sigma$ for weak lensing if marginalizing over one of the variables ($\Delta \chi^2 = 1$).
Best fit values are also indicated.
We marginalized over the weak lensing background redshift uncertainty $z_s = 0.99 \pm 0.10$.
%This figure neatly demonstrates the requirement for both strong and weak lensing data
%to precisely constrain the mass profile of this cluster.
}\end{figure}
%%%%%%%%%%%%%%%%%%%%%%%%%%%%%%%%%

%%%%%%%%%%%%%%%%%%%%%%%%%%%%%%%%%
\begin{deluxetable*}{cccl}
\tablewidth{0pt}
\tablecaption{\label{tab:Mc}Published NFW Fits to the Mass Profile of A2261}
\tablehead{
\colhead{Virial Mass}&
\colhead{Concentration}&
\colhead{}&
\colhead{}\\
%\colhead{$M_{vir} (10^{15} M_{\odot})^{a}$} &
\colhead{$M_{vir} (10^{15} M_{\odot} \hh$)} &
\colhead{$c_{vir}$}&
\colhead{article}&
\colhead{constraints}
}
\startdata
$1.93^{+0.37}_{-0.31}$ &
$6.4^{+1.9}_{-1.4}$ &
\cite{Umetsu09} &
1) WL shears\\
$1.80^{+0.29}_{-0.24}$ &
$10.2^{+7.1}_{-3.5}$ &
\cite{Umetsu09} &
2) WL shears + estimated mass sheet$^a$\\
$1.79^{+0.24}_{-0.23}$ &
$11.1^{+2.2}_{-1.9}$ &
\cite{Umetsu09} &
3) WL(2) + Einstein radius estimate:$^b$ $R_E(z_s = 1.5) = (40 \pm 2)\arcsec$\\
%$R_E = (40 \pm 4)\arcsec$ for $z_s = 1.5$\\
%... + $R_E = 40\arcsec$ ($z_s = 1.5$)\\
%WL kappa + SL $R_E(z_s = 1.5) = 40\arcsec$\\
%
$1.36^{+0.29}_{-0.24}$ &
$6.0^{+1.7}_{-1.3}$ &
\cite{Okabe10} &
4) WL shears\\
%
%$2.03^{+0.35}_{-0.28}$ &
%$6.1^{+1.1}_{-1.0}$ &
%this work &
%5) WL shears\\
%
%$1.32^{+0.29}_{-0.24}$ &  % / h
$1.89^{+0.41}_{-0.34}$ &
$5.8^{+1.8}_{-1.4}$ &
this work &
5) WL(2) re-analyzed$^c$\\
%
% Keiichi e-mail
$2.09^{+0.31}_{-0.27}$ &
$6.0^{+1.1}_{-0.9}$ &
this work &
6) WL shears\\
%6) WL shears$^d$\\
%
% python cMrange.py -2 61 W
%$1.90^{+0.25}_{-0.25}$ &
%$6.6^{+1.4}_{-1.0}$ &
$1.89^{+0.25}_{-0.22}$ &
$6.7^{+1.1}_{-1.0}$ &
this work &
7) WL shears + estimated mass sheet$^d$\\
%7) WL shears + estimated mass sheet$^e$\\
%
% python cMrange.py -3 121 W
%$2.23^{+0.29}_{-0.26}$ &
%$5.6^{+0.9}_{-0.8}$ &
%$2.25^{+0.99}_{-0.68}$ &
%$5.7^{+1.0}_{-0.7}$ &
$2.21^{+0.25}_{-0.23}$ &
$5.7^{+0.8}_{-0.7}$ &
this work &
8) WL shears + magnification (number count depletion)\\
$1.97^{+0.26}_{-0.21}$ &
$6.6^{+0.5}_{-0.4}$ &
this work &
9) SL + WL(5)\\
%
% python cMrange.py -2 61
% ~/A2261/NFW/new/notes.txt
%$1.97^{+0.20}_{-0.20}$ &
%$6.6^{+0.5}_{-0.4}$ &
$1.98^{+0.19}_{-0.16}$ &
$6.6^{+0.4}_{-0.4}$ &
this work &
10) SL + WL(7)\\
$\Mvirsphd$ &
$\cvirsphd$ &
this work &
11) SL + WL(8) = our primary result when assuming a spherical halo\\
%
%$1.79^{+0.18}_{-0.14}$ &
%$4.4^{+0.2}_{-0.3}$ &
%$1.65^{+0.16}_{-0.12}$ &
%$4.7^{+0.2}_{-0.3}$ &
$\Mvirelld$ &
$\cvirelld$ &
this work &
12) SL + WL(8) + X-ray, with one model for halo elongation\\
%11) SL + WL(7) + X-ray, allowing for halo elongation\\
%
\vspace{-0.1in}
\enddata
\tablecomments{We also roughly identify and estimate the effects of 
background/foreground structures along the line of sight (\S\ref{sec:LSS}).
We find that correcting for these may lower $M_{vir}$ by $\sim 7\%$
and increase $c_{vir}$ by $\sim 5\%$.}
%\tablenotetext{a}{Assuming $h=0.7$.}
\tablenotetext{a}{The mass sheet density $\kappa_b$ in the outer annulus was estimated based on an NFW fit to the shears.}
\tablenotetext{b}{In this work, we find a lower $R_{E} (z_{s} = 1.5) = (20 \pm 2) \arcsec$.}
%\tablenotetext{c}{Now including innermost profile point, plus improved shear calibration 
%which increases $M_{vir}$ by $\sim 5\%$ as expected.}
%\tablenotetext{c}{Now including the central bin $\bar{\kappa}(<\theta_{min})$, the mean convergence interior to the inner radial boundary of weak lensing measurements,  $\theta_{min}=1\arcmin$.}
%\tablenotetext{c}{Now including the innermost bin $\bar{\kappa}(< 1')$, 
%the mean convergence interior to the weak lensing measurements.}
\tablenotetext{c}{Now including the innermost bin $\bar{\kappa}(< 1')$, 
the mean convergence interior to the weak lensing measurements.}
%as well as improved shear calibration.}
%\tablenotetext{d}{Including improved shear calibration which increases $M_{vir}$ by $\sim 5\%$, as expected.}
%\tablenotetext{d}{Including improved shear calibration which increases WL signals by $\sim 5\%$.}
%\tablenotetext{d}{Including improved shear calibration which contributes a $\sim 5\%$ increase to $M_{vir}$.}
\tablenotetext{d}{Iterative NFW fitting is performed to find the best fitting mass sheet density $\kappa_b$.}
%to $\kappa(R)$
\end{deluxetable*}
%%%%%%%%%%%%%%%%%%%%%%%%%%%%%%%%%

\section{Mass Profile from Joint Strong + Weak Lensing Analysis}
\label{sec:lensing}

%For a more robust and precise determination of the mass concentration,
For a mass concentration determination that is both precise and accurate,
the inner mass profile must be simultaneously constrained by strong and weak lensing analyses
\citep[e.g.,][]{Meneghetti10}.
In \S\ref{sec:SL} we derived seven SL mass profiles from which we calculated the average with uncertainties.
Then in \S\ref{sec:WL} we presented various WL analyses.
Our final WL analysis including both shear and magnification (number count depletion) information is our most robust.
The magnification data break the mass sheet degeneracy and increase our overall WL signal-to-noise.

Our strong and weak lensing data agree well in their region of overlap (Fig.~\ref{fig:massprofileprojected}).
We perform joint NFW fitting to
the SL mass enclosed $M(<R)$ measured at 
%12 points between $5''$--$1'$
12 points $5'' \leq R \leq 1'$ (18 -- 215 kpc)
and the WL mass density $\kappa(R)$ measured in 11 bins with centers 
$40'' \leq R \lesssim 14.2'$ (144 -- 3,059 kpc).
%between $40''$ and $14.2'$.

This yields a virial mass \Mres\ and concentration \cres\ with a 
significantly greater precision than that obtained by WL alone ($c_{vir} = 5.7^{+1.0}_{-0.7}$).
Confidence contours are plotted in Fig.~\ref{fig:M-c}
and the constraints are tabulated in Table \ref{tab:Mc}.
Our new results strongly disfavor the previous $c_{vir} \sim 10$ results \citep{Umetsu09}.

Our use of the spherical NFW profile enables the most direct comparisons
with analyses of simulated halos fit exclusively to this profile \citep[e.g.,][]{Klypin11,Prada11,Bhattacharya12}.
Other mass profiles, including the \cite{Einasto65} profile, 
have been shown to yield slightly better fits to simulated halos \citep{Navarro04,Merritt05,Merritt06,Navarro10}.
The choice of profile does not significantly affect the derived concentrations \citep{Duffy08,Gao08,Reed11}.
%We will explore this further with our observational data in future work.

%%%%%%%%%%%%%%%%%%%%%%%%%%%%%%%%%
\begin{figure*}
\epsscale{1.18}
%\plotone{figs/SubaruWL.jpg}
%\plotone{figs/SubaruWLpeaks.jpg}
%\plotone{figs/A2261_peakz600.png}
%\plottwo{figs/A2261_peakz600.png}{figs/WLmassmapholes.png}
%\begin{center}
%  \includegraphics[width = .42\textwidth]{figs/WLcolorlegend75sq.png}
%  \includegraphics[width = .56\textwidth]{figs/WLmassmapholes.png}
%\end{center}
\plotone{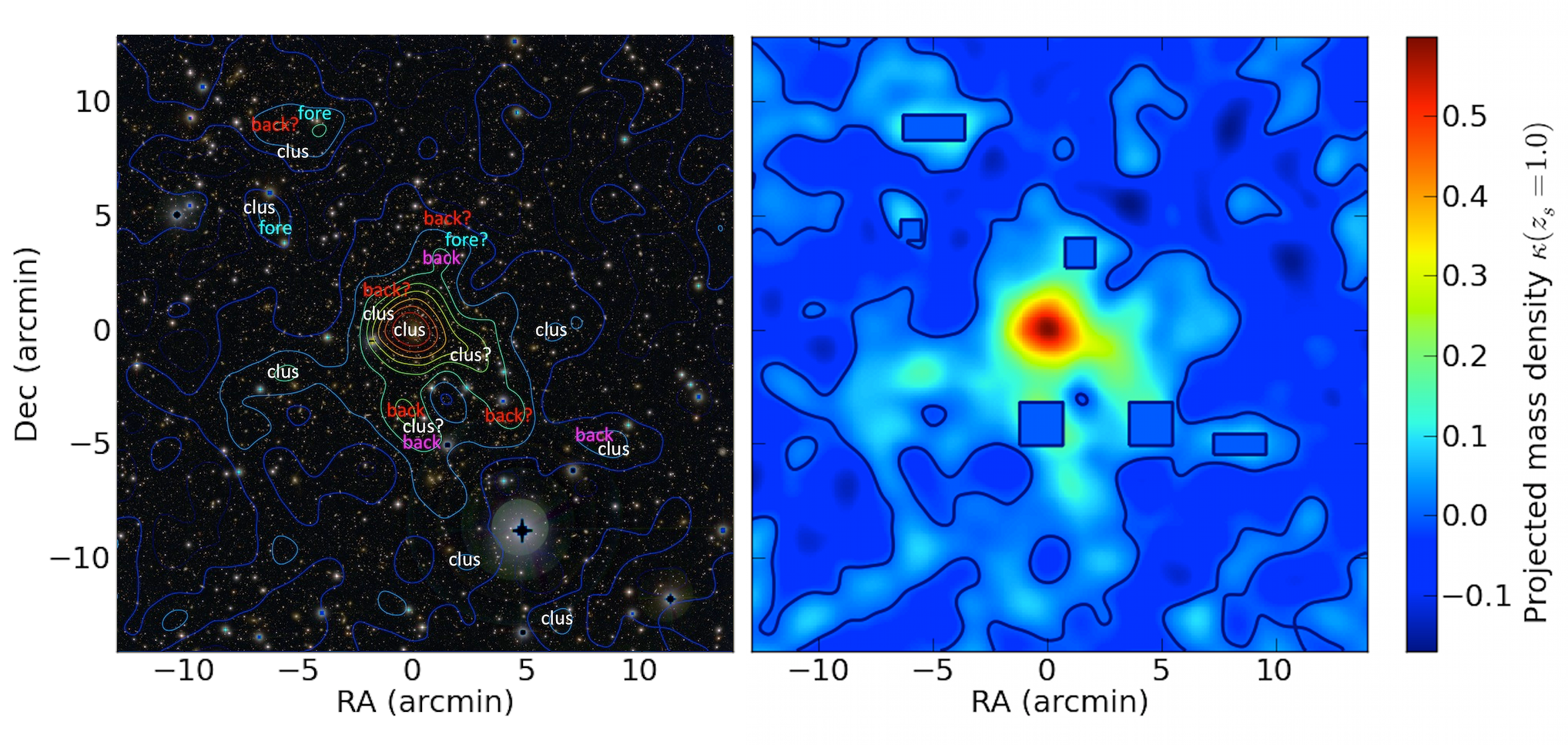}
\caption{\label{fig:SubaruWL}
Weak lensing mass reconstruction of A2261 based on analysis of Subaru images.
{\emph Left}: Mass contours are overlaid on this Subaru \BVR\ color image
$27\arcmin \sim 5.8$ Mpc on a side.
Mass peaks are tentatively identified as belonging either to the cluster or background/foreground structures
based on nearby galaxies with spectroscopic redshifts from Hectospec
or 5-band Bayesian photometric redshifts (labeled with question marks ``?'').
Magenta labels correspond to background galaxies at $0.25 < z < 0.28$
with recession velocities 7,500 -- 16,500 km$/$s greater than the cluster.
Red labels correspond to $0.40 < z < 0.53$.
\emph{Right}: Mass peaks are somewhat aggressively set to zero
where contributions from background/foreground structures are suspected.
The dark contour corresponds to zero projected overdensity.
This is used to estimate the effects of large scale structure on our mass profile.
We note this linear \cite{KaiserSquires93} reconstruction was smoothed with a Gaussian kernel
and not used in our analysis aside from this estimate.
}\end{figure*}
%%%%%%%%%%%%%%%%%%%%%%%%%%%%%%%%%

\subsection{Effect of Background Structures}
\label{sec:LSS}

Significant background structure was identified by \cite{Umetsu09} and \cite{Okabe10}
and estimated to be at $z \sim 0.5$ based on $\VJm-\RCm$ galaxy colors in that region.
It was posited that this structure could bias the derived weak lensing signals.
Here we present a rough estimate of the effects of background structures
on our derived mass and concentration.

We identified mass peaks in a weak lensing mass model
obtained using a linear \cite{KaiserSquires93} mass reconstruction method with Gaussian smoothing 
(Fig.~\ref{fig:SubaruWL}).
We then estimated redshifts for the twelve peaks with nearby bright galaxies
based on spectroscopic and photometric redshift information.
Hectospec spectroscopic redshifts were available for ten of the peaks.
For the remaining galaxies, we used BPZ photometric redshifts derived using their \BVRiz\ magnitudes.
We note these achieved a good accuracy of $\sim 3\% (1+z)$
for the $\sim 300$ galaxies with \BVRiz\ photometry and confident spectroscopic redshifts.
% (\S\ref{sec:widefield}).

We identified six mass peaks coincident with bright galaxies in the background or foreground.
We then eliminated those peaks from our mass model %somewhat aggressively
by setting the overdensity of those regions equal to zero (Fig.~\ref{fig:SubaruWL})
and rederived the mass profile as determined by weak lensing.
Based on fitting of NFW profiles to our strong and weak lensing,
we found that removal of these background structures
lowered the virial mass $M_{vir}$ by $\sim 7\%$
and increased the concentration $c_{vir}$ by $\sim 5\%$.
(Fitting to WL alone yielded slightly higher $\sim 10\%$ effects.)
%
%Though the estimate is rough,
We conclude that background structures likely affect the mass and concentration measurements 
from joint SL+WL fitting at the 10\% level or less.
We made some attempt to maximize this effect
by setting mass overdensities equal to zero
(some overdensity should remain in these regions due to the cluster).
However our analysis was not extreme either in the number or sizes of areas eliminated.
% or numbers of background mass peaks identified.
%To be even more aggressive we could have modified larger regions and/or identified additional background mass peaks.
%While setting projected mass overdensities equal to zero is aggressive
%(some overdensity should remain due to the cluster),
%our box sizes were not necessarily aggressive.
%Rather than setting the projected mass overdensity equal to zero within the squares,
%we may have allowed them some finite value consistent with the cluster density profile.
%However we may have also made the squares larger to more fully eliminate the targeted mass peaks.

%===================================================

\section{Triaxiality from Joint Lensing + X-ray Analysis}
\label{sec:joint}

Lensing analysis may yield higher mass estimates than X-ray analysis
for either or both of the following reasons:
1) halo elongation and/or additional massive structures along the line of sight boosting the lensing signal
\citep{Meneghetti10,Newman11,MorandiLimousin11A383};
2) non-thermal gas pressure support 
(primarily turbulent flows and/or bulk motions)
% + cosmic rays + magnetic fields??
% http://iopscience.iop.org/0004-637X/675/1/126/fulltext
%absent from the X-ray signal and 
deviating from assumptions of hydrostatic equilibrium
\citep{Nagai07,Lau09,Kawaharada10}.

In cosmological simulations, dark matter halos are generally found to be prolate
with typical axis ratios of $\sim$ 2:1.
This elongation is generally found to decrease as a function of radius
\citep{BarnesEfstathiou87,Warren92,JingSuto02,Schulz05,Lemze11}.
% \bibitem[Lemze et al.(2011)]{2011arXiv1106.6048L} Lemze, D., Wagner, R., Rephaeli, Y., et al.\ 2011, arXiv:1106.6048 
This trend may be dampened by baryons
which are more dominant at smaller radii
and act to make halos more spherical due to their collisional nature \citep[e.g.,][]{Kazantzidis04}.
Halo elongations along the line of sight 
can bias both lensing strengths and cluster concentration measurements significantly high,
such that the measured concentrations of a lensing-selected sample may be biased high by $\sim 50$-100\%
\citep{Hennawi07,OguriBlandford09,Meneghetti10,Meneghetti11}.

Non-thermal pressures may account for $\sim 15\%$ of the total support against gravitational collapse,
thus biasing low by that amount the mass derived when assuming hydrostatic equilibrium 
\citep{Nagai07,Lau09}.
In relaxed clusters,
non-thermal pressure support is expected
%to be greatest in the outskirts due to inflowing gas
to increase with radius up to $\sim 30$-40\% at the virial radius due to inflowing gas
(\citealt{Lau09,Shaw10b,Cavaliere11a}).
A possible minimum in the non-thermal pressure support
at $\sim 0.1 r_{vir}$ has also been predicted \citep{Molnar10}.
%\citep{Cavaliere11a}.
%to increase as a function of radius
%due to inflows at the outskirts \citep{Cavaliere11a}.

Previous joint lensing + X-ray analyses have allowed for these factors as global constants 
\citep[e.g.,][]{Morandi10,Newman11,Morandi11,MorandiLimousin11A383}.
Radial dependence of non-thermal pressure support was modeled by \cite{Morandi11c}.
Here we consider radial variation of this quantity as well as elongation.
We only consider elongation along the line of sight
as our 2D lens mass modeling already allows for elongation and more general asymmetries within the plane of the sky.

%\subsection{Observations: Chandra X-ray Imaging}
\subsection{Chandra X-ray Observations and Analysis}

% 9.18, 24.64 ksec according to Chandra archive
A2261 was observed by Chandra ACIS-I in programs \#550 and \#5007 (P.I. Van Speybroeck)
to depths of 9.0 and 24.3 ksec, respectively 
\citep{Morandi07,Maughan08,Gilmour09,Mantz10b}.

We reprocessed and filtered the X-ray events in the latter observation in a standard manner 
using CIAO v4.3 and CALDB version 4.4.6. 
Based on $\sim 24,000$ net photon counts (0.7--7.0 keV),
we extracted X-ray spectra 
within 15 annuli in the range $6'' < R < 3.1'$
(20 kpc $\lesssim R \lesssim 650$ kpc)
%($\sim 20$ -- 650 kpc)
centered on the X-ray peak which is coincident with the center of the BCG.
There were roughly equal net counts per annulus.
%There were roughly equal numbers of photons per annulus in the 0.7--7.0 keV passband.
A matched extraction of events from a reprojected, filtered, deep background events file 
%(Markevitch ACIS memo) 
was used for the background spectrum.

XMM observed Abell 2261 on 9 separate occasions between 2003 and 2004 for $\sim$12--13 ksec each. 
Each observation was heavily contaminated by proton flares and deemed unsuitable for analysis. 
It is likely that these lower priority observations were scheduled during
periods of elevated particle backgrounds.

We fit the Chandra spectra simultaneously by creating models of hot gas in hydrostatic equilibrium in a dark matter 
NFW gravitational potential well using the JACO (Joint Analysis of Cluster Observations) software \citep{Mahdavi07a}.
%(Mahdavi et al. 2007; etc)
JACO allows for nuisance parameters such as an X-ray point source (none was detected) 
and contributions from a galactic soft background (found to be negligible in this case).
%[A little blurb about JACO, gory detail are in the two main papers...]
%[ I did one fit where the full mass profile follows NFW, another where only the dark matter is assumed to follow NFW, the gas is assumed in hydrostatic equilibrium with the dark matter, but can have a core.]

\begin{figure}
\epsscale{1.2}
%\plotone{figs/masswithin.pdf}
%\caption{\label{fig:masswithin}
%\plotone{figs/A2261massprofile.png}
%\plottwo{figs/A2261massprofile22.png}{figs/A2261massprofile23.png}
%\plottwo{figs/A2261massprofile35.png}{figs/A2261massprofile39.png}
%\plotone{figs/A2261massprofile39.png}
%\plotone{figs/A2261massprofile39.pdf}
%\plotone{figs/A2261massprofile.pdf}
\plotone{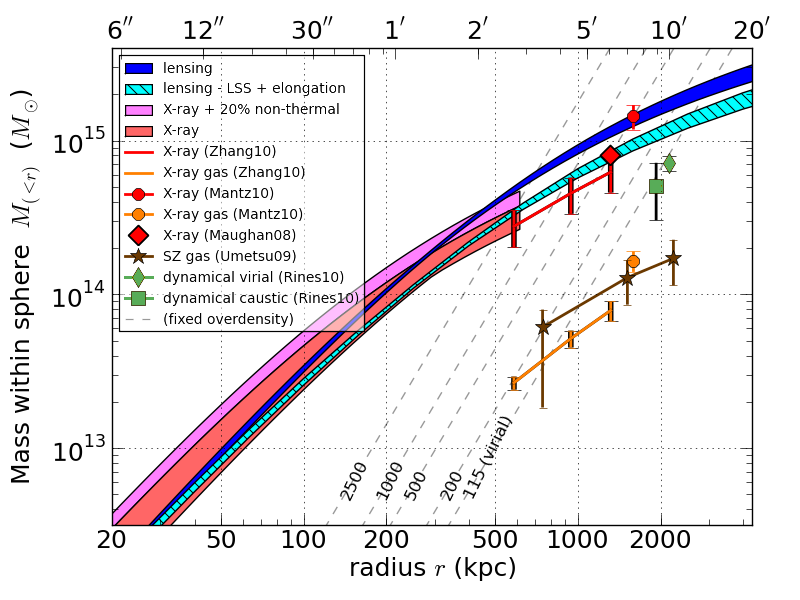}
\caption{\label{fig:massprofile}
%Mass profiles derived from lensing, X-ray, and dynamical analyses,.
Mass profiles derived from various observational probes.
We derive an X-ray mass (red curve, NFW profile) 
$\sim 35\%$ lower than our lensing mass (blue, NFW profile) at $r_{2500} \sim 600$ kpc.
Marginal agreement can be achieved by allowing for 20\% non-thermal pressure support (magenta),
though this is much higher than generally expected at this radius \citep[e.g.,][]{Lau09}.
Agreement may be more readily achieved by an elongated halo
with a 2:1 axis ratio for $r > 100$ kpc (light blue hashed).
However, the need for such elongation may be obviated completely
by systematic uncertainties in the X-ray results \citep{Maughan08,Zhang10,Mantz10b}.
The latter result agrees well with our spherical lensing mass profile.
A similar range of results was found by \cite{Conte11} who consider various systematics.
The dynamical mass estimates \cite{Rines10} are significantly lower 
($M_{100} \sim 0.5$ -- $0.7 \times 10^{15} M_\odot h_{70}^{-1}$).
%These will be analyzed and discussed further by Lemze et al.~2012, in preparation.
Also plotted are gas mass measurements based on X-ray (orange) 
and AMiBA SZE observations (brown stars; \citealt{Umetsu09}).
}\end{figure}

In Fig.~\ref{fig:massprofile}, we plot our NFW fit to the total mass (gas + dark matter) profile
assuming a spherical halo and hydrostatic equilibrium (HSE).
We fit out to $r = 3.1' \approx 667$ kpc $\hh$,
or just beyond $r_{2500} = 590$ kpc $\hh$,
corresponding to $M_{2500} =  (0.29 \pm 0.05) \times 10^{15} M_\odot \hh$,
which we derive along
with an NFW concentration $c_{2500} = 2.3 \pm 0.9$.
This mass is $\sim 35\%$ lower than the mass we derive at that radius based on our lensing analysis.
%and a gas fraction $f_g = XX \pm XX h_{70}^{-3/2}$.
For reference,
if extrapolated to the virial radius, this profile would correspond to
$M_{vir} = (0.82 \pm 0.14) \times 10^{15} M_\odot \hh$
with
$c_{vir} = 9.1 \pm 3.0$.
As we show, this is in good agreement with \cite{Zhang10} who also fit out to larger radii using the XMM data.
\cite{Maughan08} find a slightly larger $M_{500} \sim 0.80 \times 10^{15} M_\odot$
within $R_{500} \approx 1.31$ Mpc
based on the Chandra data.

We also plot 20\% deviations from HSE in the form of non-thermal pressure support.
Though larger than expected within $r_{2500}$
\citep{Lau09,Shaw10b,Molnar10,Cavaliere11a,Nelson11},
this is what the data would require to bring the lensing and X-ray masses derived in this work
into agreement just within the error bars.

%\cite{Mantz10b} and \cite{Umetsu09} derive higher gas masses than \cite{Zhang10}
%based on their Chandra X-ray and AMiBA SZE analyses, respectively
%(also shown in Fig.~\ref{fig:massprofile}).
\cite{Mantz10b} derive a higher gas mass than \cite{Zhang10}
(also shown in Fig.~\ref{fig:massprofile}).
Based on this, they derive a significantly higher $M_{500} = 1.44 \pm 0.26 \xfm$
within $r_{500} = 1.59 \pm 0.09$ Mpc.
This is in excellent agreement with our derived lensing mass.

\cite{Mantz10b} assume a gas mass fraction $f_{gas} \sim 12\%$ for A2261,
very similar to the $f_{gas}$ derived by \cite{Zhang10} assuming hydrostatic equilibrium.
Various systematics are discussed further in \cite{Conte11}
who also derive a range of mass estimates for A2261 similar to that described already.
We consider this full range in our analysis.

We note that published dynamical mass estimates of A2261 are significantly lower \citep{Rines10}.
These data are somewhat limited by bright stars in this field,
hindering our ability to obtain additional spectra which might resolve this discrepancy.

X-ray observables, and the masses derived from them, are largely insensitive to halo elongation
\citep[e.g.,][]{Gavazzi05,Nagai07,Buote11a,Buote11b}.
This is not the case for masses derived from lensing data, as we discuss below.

%%%%%%%%%%%%%%%%%%%%%%%%%%%%%%%%%
%%%%%%%%%%%%%%%%%%%%%%%%%%%%%%%%%

%\subsection{Halo Elongation and Non-Thermal Pressure Support}
\subsection{Halo Elongation}
\label{sec:elongation}

As shown in Fig.~\ref{fig:massprofile},
our mass profiles derived independently from lensing and X-ray analyses are in good agreement in the core,
while the latter exhibits a $\sim 35\%$ deficit at the X-ray $r_{2500} \sim 600$ kpc.
This result is toward the low end of other X-ray mass estimates, so we consider this to be a limiting case.
This deficit could best be accounted for by halo elongation along our line of sight
(though the \citealt{Mantz10b} result would obviate the need for any such elongation).

%This deficit can best be accounted for by halo elongation along our line of sight.
We found that an axis ratio of 2:1 is able to 
bring our lensing and X-ray results into better agreement at $r_{2500}$.
This elongation is not required at inner radii where a spherical profile fits the data.
Halo elongation is generally expected to {\em decrease}, not increase, with radius \citep[e.g.,][]{Hayashi07}.
However we note that here we are probing the very inner core
where the dense concentration of baryons may increase the sphericity \citep[e.g.,][]{Kazantzidis04}.
The large BCG of A2261 extends visibly to $r \sim 100$ kpc.

We construct a toy model for the halo
elongation $e = 1 - b/a$ varying with radius,
% ($e = 1 - b/a$)
increasing from zero (spherical) for $r \leq 1$ kpc
to 0.5 (an axis ratio of 2:1) beyond $r \geq 100$ kpc.
Between these two radii, it follows $e(r) = 0.25 \log_{10}(r / {\rm kpc})$.
%(Thus, for example, $e = 0.75$ at $r = 10$ kpc.)
The 3D mass density $\rho(r)$ scales with halo roundness ($\xi = 1 - e$):
%$\rho(r) = \xi {\rm NFW}(u)$
$\rho(r) = \xi(r) \rho_{\rm NFW}(u)$
where
$u = \sqrt{x^2 + y^2 + {\xi z}^2}$
and $\rho_{NFW}(u) = \rho_{s} (u / r_{s})^{-1} (1 + u/r_{s})^{-2}$.
This scaling preserves the projected mass density $\kappa(R)$ integrated along the line of sight ($z$-axis)
and thus preserves all lensing observables.

We applied this elongation profile to our
primary joint (SL + WL shear + magnification)
lensing profile NFW fit: \Mres\ and \cres.
We then calculated numerically the 3D mass enclosed
within spherical shells for this ellipsoidal mass distribution.
This is plotted as the light blue hashed region in Fig.~\ref{fig:massprofile}.
We find this model agrees well with 
both our lensing and X-ray derived mass profiles,
whether including modest non-thermal pressure support or not.
%We leave for future work the exploration of a broader range of elongation profiles.

%We derive a new NFW fit for our elongated profile:
A spherical NFW fit to this elongated profile yields \Mell\ and \cell.
%$M_{vir} = 1.65^{+0.16}_{-0.12} \times 10^{15} M_{\odot}$
%and $c_{vir} = 4.7^{+0.2}_{-0.3}$.
%$M_{vir} = 1.65^{+0.16}_{-0.12} \times 10^{15} M_{\odot}$
%and $c_{vir} = 4.7^{+0.2}_{-0.3}$.
%This is given in Table \ref{tab:Mc} and plotted in 
%as the light blue hashed lin in Fig.~\ref{fig:massprofile}.
%
Applying the corrections for background / foreground line of sight structures estimated in \S\ref{sec:LSS},
we find $M_{vir} \sim 1.6 \times 10^{15} M_{\odot}$ (a $\sim 7\%$ decrease)
and $c_{vir} \sim 4.8$ (a $\sim 5\%$ increase).
Note the former corrections for cluster halo elongation
are significantly larger than those for LOS structure.

% FROM NEWMAN10
%"Finally, we recall that the A08 X-ray measurements assumed sphericity, whereas our mass models are non-spherical. By calculating the gas emission in a non-spherical halo with qDM = 2, we estimate that spherically deprojecting the X-ray observables biases the inferred (spherically averaged) mass profile by only 7% typically, consistent with previous studies (Gavazzi 2005; Nagai et al. 2007). As discussed in Section 2.4, this small bias is comparable to other systematic uncertainties inherent to X-ray analyses and is within our adopted calibration uncertainty. We estimate the impact on our results by shifting the X-ray masses accordingly."

%\subsection{Elongated Concentration}
%\label{sec:elongation}

%$M_{vir}$ decreases $\sim 7\%$ to $\sim 1.5 \times 10^{15} M_{\odot}$,
%and $c_{vir}$ rises $\sim 5\%$ to $\sim 4.9$.
% ~/c-M/ellNFW/ellvarNFW.py -> ellvarNFW33.dat
%1.65103e+15  2721.45  4.68314  1.22628
% Mfac = 0.931226765799
% cfac =  1.05229872415

\begin{figure}
\epsscale{1.2}
%\plotone{figs/c-M_A2261.pdf}
%\plotone{figs/cM.png}
%\plotone{figs/cMrel13.png}
%\plottwo{figs/cMrel13.png}{figs/cMall2.png}
%\plotone{figs/cMrel0.png}
%\plotone{figs/cMrel.pdf}  % doesn't capture hatching
\plotone{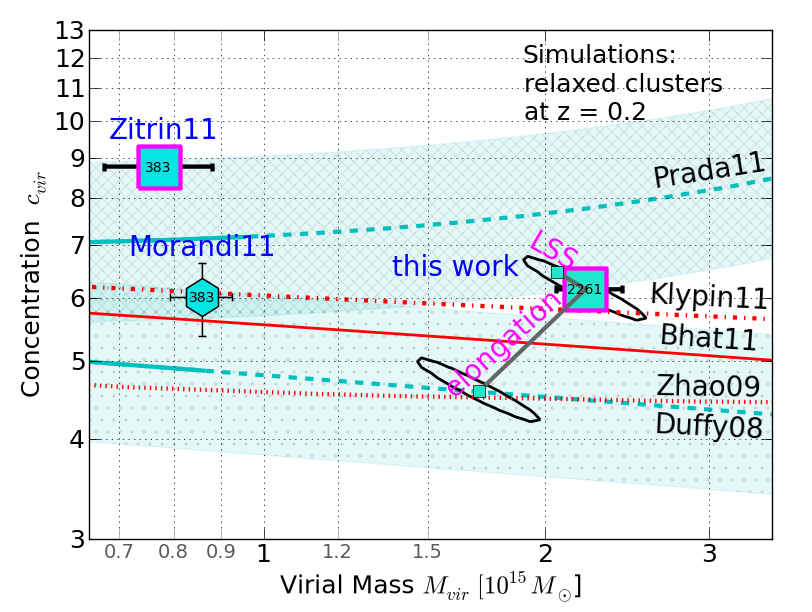}
\caption{\label{fig:c-Mrel}
Observed virial masses $M_{vir}$ and concentrations $c_{vir}$ for CLASH clusters
compared to the average $c(M,z)$ realized for relaxed clusters in simulations.
%\citep{Duffy08,Zhao09,Klypin11,Prada11, Bhattacharya12}.
Squares are from joint strong + weak lensing analyses of A2261 (this work) and A383 \citep{Zitrin11A383}.
The hexagon is from 
\citet[and private communication]{MorandiLimousin11A383} 
who fit triaxial halos to A383 SL + X-ray data.
For A2261, we plot both error bars (1-$\sigma$, marginalizing over the other parameter)
and confidence contours (1-$\sigma$).
%The error budget for A2261 is much larger as additional factors are included.
Systematic uncertainties are labeled: 
possible halo elongation (\S\ref{sec:elongation}) and line of sight structures (\S\ref{sec:LSS}).
Results realized in two simulations \citep{Duffy08,Prada11} are shown in light blue, 
including scatters of $\sim 0.1$ in $\log_{10}(c)$ ($\sim 26\%$).
Portions of these lines are dashed to indicate extrapolations to high masses
where clusters are not realized in sufficient numbers.
Averages results from three additional simulations 
\citep{Zhao09,Klypin11, Bhattacharya12}
are shown in red with styles solely for clarity.
Results are plotted for relaxed cluster subsamples as determined by \cite{Duffy08} and \cite{Bhattacharya12},
yielding concentrations $\sim 10\%$ higher than for the full populations.
This 10\% factor is applied to the results from the other simulations.
}\end{figure}

We note that this measurement method 
is consistent with that generally used to measure the mass profiles of simulated clusters.
Enclosed mass (or, more often, density) is determined assuming spherical symmetry
(and most often fit to an NFW profile)
even though the halos are triaxial and asymmetric.
%Other formulations, such as a triaxial potential, yield unphysical mass distributions for moderate ellipticities $e \gtrsim 0.3$
%\citep{LeeSuto03}.
%
Another approach is to fit the lensing and X-ray observables to an ellipsoidal NFW profile,
as in the \cite{MorandiLimousin11A383} analysis of Abell 383, the first observed CLASH cluster.
Notably, they allow for 
a fully general ellipsoidal gNFW (generalized NFW with variable inner slope) dark matter halo
plus an exponential ICM profile including non-thermal pressure support.
Ideally, simulations will be analyzed in the same way allowing for direct comparisons.
Until then, the advantages of this parameterization will not be completely realized,
as spherical averages must be derived for comparison with most published analyses of simulations.
\cite{MorandiLimousin11A383} derive $M_{vir} = 8.6\pm0.7 \times 10^{14} M_{\odot}$ and $c_{vir} = 6.0 \pm 0.6$
(private communication) based on a joint SL + X-ray analysis.
We compare this to the
%$M_{vir} = 6.3 \pm 0.7 \times 10^{14} M_{\odot}$ and
%$c_{vir} = 8.6 \pm 0.2 ({\rm stat.}) \pm 0.4 ({\rm syst.})$ 
$M_{vir} = 7.7 \pm 1.0 ({\rm stat.}) \pm 0.4 ({\rm syst.}) \times 10^{14} M_{\odot}$ and
$c_{vir} = 8.8 \pm 0.4 ({\rm stat.}) \pm 0.2 ({\rm syst.})$ 
found by \cite{Zitrin11A383}
who fit a spherical NFW profile to joint SL + WL constraints.
The effect of correcting for elongation is to decrease 
the derived concentration
%both derived parameters,
as in our analysis of A2261
(see Fig.~\ref{fig:c-Mrel}).

%$4.8 \pm 0.5$
%$8.6 \pm 0.2$ (stat.) $\pm 0.4$ (syst.).
%dark matter + ICM,
%with the former being a fully general gNFW ellipsoidal dark matter halo
% - X-ray temperature (hydrostatic equilibrium component, so depends on non-thermal fraction)
% - X-ray surface brightness (doesn't depend on non-thermal fraction)

%XXX Discussion: Morandi11 -- Just SL + X-ray?  Parameterization...

%%%%%%%%%%%%%%%%%%%%%%%%%%%%%%%%%
%\begin{figure}
%\epsscale{1.1}
%\plotone{figs/REzs_Adi_Dan_NFW.pdf}
%\caption{\label{fig:RE}
%Einstein radius as a function of redshift.
%}\end{figure}
%%%%%%%%%%%%%%%%%%%%%%%%%%%%%%%%%

%%%%%%%%%%%%%%%%%%%%%%%%%%%%%%%%%
%%%%%%%%%%%%%%%%%%%%%%%%%%%%%%%%%

%%%%%%%%%%%%%%%%%%%%%%%%%%%%%%%%%
\begin{figure}
\epsscale{1.2}
%\plotone{figs/c-M_A2261.pdf}
%\plotone{figs/cM.png}
%\plotone{figs/cMrel13.png}
%\plottwo{figs/cMrel13.png}{figs/cMall2.png}
%\plotone{figs/cMall2.png}
\plotone{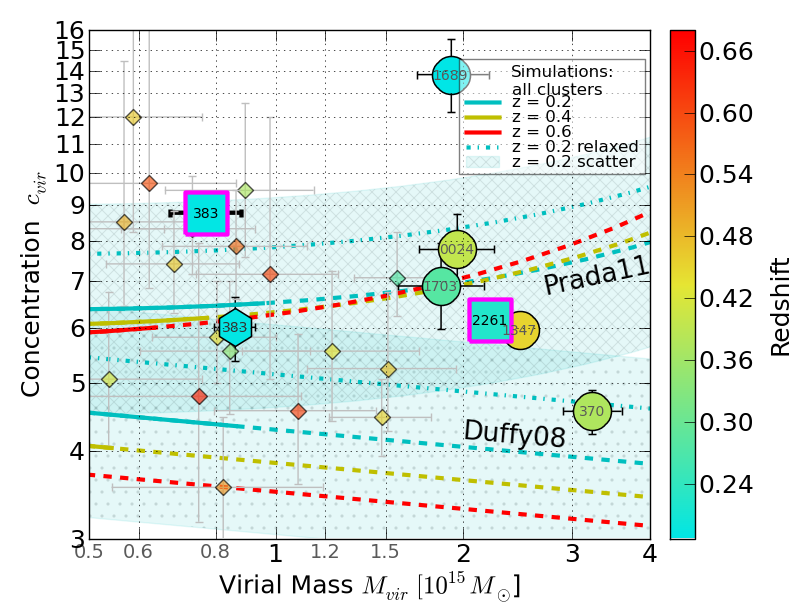}
\caption{\label{fig:c-Mall}
Similar to Fig.~\ref{fig:c-Mrel} but now for all clusters (not just relaxed).
The additional data points are all analyses of non-CLASH, lensing-biased clusters,
as follows and all colored accoring to redshift.
Circles are from \cite{Umetsu11} SL+WL analyses.
And small diamonds are from \cite{Oguri11} analyses
with WL + basic SL constraints (the Einstein radii).
%Note the average predictions of $c(M,z)$ from simulations are $\sim 25\%$ lower for all clusters than for relaxed clusters.
Note the average predictions from simulations for $c(M,z)$ for all clusters
%are $\sim 25\%$ lower than for relaxed clusters.
are $\sim 10\%$ lower than for relaxed clusters.
The expected scatters are larger: $\sim 0.15$ in $\log_{10}(c)$ ($\sim 41\%$).
}\end{figure}
%%%%%%%%%%%%%%%%%%%%%%%%%%%%%%%%%

\section{Mass Profile Compared to Simulated Clusters}
\label{sec:c-M}

Based on our joint strong + weak lensing + X-ray analysis,
we find that A2261 is \emph{not} significantly over-concentrated 
compared to simulated relaxed clusters of similar mass and redshift.
This is demonstrated in Fig.~\ref{fig:c-Mrel}.
Our range of allowed concentrations ($4.4 \lesssim c \lesssim 6.5$)
span the low end of average expectations ($4.5 \lesssim \langle c \rangle \lesssim 7.8$)
from simulations \citep{Duffy08,Zhao09,Klypin11,Prada11,Bhattacharya12}.
Results are also plotted for the first CLASH cluster Abell 383 \citep{Zitrin11A383, MorandiLimousin11A383}.

Note the recent Bolshoi and Multidark simulations \citep{Klypin11,Prada11}
produce halos with significantly higher concentrations 
%(by $\sim 50\%$ on cluster scales, according to the latter)
than previous simulations 
\citep[although see \citealt{Hennawi07}]{Neto07,Maccio08,Duffy08,Zhao09}.
\cite{Prada11} find upturns in $c(M,z)$ for high masses and redshifts.
This behavior is unexpected and its origin needs to be understood.
\cite{Bhattacharya12} find no evidence for such upturns in their analysis of an even larger simulation,
2 Gpc $h^{-1}$ on a side, 8 times the volume of Multidark, with the same number of particles ($2048^3$).
%The \cite{Prada11} results are especially intriguing
%because they find upturns in $c(M,z)$ for high masses and redshifts.
%This behavior is unexpected and its origin has yet to be determined.
%This origin of this unexpected behavior has yet to be determined.
%This origin of this disagreement has yet to be determined.

A383 and A2261 are two of 20 CLASH clusters selected based on X-ray properties.
We expect this sample to be less biased toward elongations along the line of sight
than a lensing-selected sample.
However some bias may remain.
%Perhaps clusters which are roughly round and symmetric in the X-rays are preferentially elongated along the line of sight.
If we assume all clusters are prolate and elongated in some direction,
perhaps clusters which are roughly round and symmetric in the X-rays may preferentially be elongated along our line of sight.
%We will continue to measure and quantify the elongations of our clusters as we have done in this paper.
We will continue to make precision measurements of the mass profiles 
and constrain DM elongation for the CLASH clusters as the survey progresses.

A2261 is borderline relaxed.
\cite{Gilmour09} classified it as disturbed,
but the X-ray peak is well aligned with the BCG,
and the X-ray luminosity is symmetric except for a subclump to the SW.
\cite{Maughan08} 
measured centroids in various annuli
and found the RMS shift to be modest
%($\langle w \rangle \sim 0.007 ~R_{500}$)
$\langle w \rangle = (7.1 \pm 0.6) \times 10^{-3} R_{500}$,
consistent with that found for relaxed clusters $\langle w \rangle \lesssim 0.012 R_{500}$.
%along with the ellipticity $e = 0.10\pm0.01$.
%
%Based on X-ray observations, it has been classified as relaxed, 
%including a small offset between the BCG and X-ray peak
%as well as low asymmetry except for a subclump to the SW.

In Fig.~\ref{fig:c-Mall}, we plot the expected $c(M,z)$ for \emph{all} clusters 
versus the most robust results from other strong + weak lensing analyses to date,
including those just mentioned plus \cite{Umetsu11} and \cite{Oguri11}.
These clusters were initially selected for study based on their lensing strength,
thus their concentrations are expected to be biased significantly high.
Disparity in this comparison is further increased as the expectations from simulations are lower.
Average concentrations for relaxed clusters (as plotted in the previous figure \ref{fig:c-Mrel})
are found to be $\sim 10\%$ higher (and have lower scatter) 
than averages for the general population as plotted in this figure \ref{fig:c-Mall}.

%%%%%%%%%%%%%%%%%%%%%%%%%%%%%%%%%
%\begin{figure}
%\epsscale{1.2}
%\plotone{figs/critcurves.png}
%\caption{\label{fig:critcurves}
%Critical curves
%}\end{figure}
%%%%%%%%%%%%%%%%%%%%%%%%%%%%%%%%%

%\section{Discussion}
%\section{Discussion: The Formation Time of A2261 from Various Observational Probes}
\section{Discussion: The Formation Time of A2261}
\label{sec:probes}

We have found the mass profile and concentration of A2261
to be in agreement with values realized in cosmological simulations for similar clusters.
This is contrary to the previous finding of \cite{Umetsu09} based solely on ground-based data
which found a high concentration suggesting an early formation time.

We can attempt to quantify this statement based on the relation
$c_{obs} \approx c_{1} a_{obs} / a_{f}$
found in previous work
\citep{Bullock01,Wechsler02,Zhao03a,Wechsler06}.
Here $a = (1 + z) ^{-1}$ is the cosmic scale factor.
The halo concentration is imprinted with the background density at its formation time
via $c(z_f) \approx c_{1} (1 + z_f)$
and then increases over time roughly as $c(z) \propto a = (1 + z) ^{-1}$.
%until we observe it as $z_{obs}$.
%$c_{vir} = c_{1} (1 + z_{f}) / (1 + z_{obs})$

The constant $c_1$ depends on the criteria used to define ``formation time''.
Various definitions have been proposed
based on the slowing rate of mass accretion 
\citep[e.g.,][]{Wechsler02,Cavaliere11b}
or mass attaining some fraction of the observed mass \citep[e.g.,][]{SadehRephaeli08}.

%%%%%%%%%%%%%%%%%%%%%%%%%%%%%%%%%
\begin{deluxetable}{cccc}
%\tablewidth{0pt}
\tablewidth{0.8\columnwidth}
\tablecaption{\label{tab:obs}A2261 Formation Redshift $z_f$ Estimates\\
Based on Various Criteria for ``Formation''}
\tablehead{
\multicolumn{3}{c}{Redshift $z_f$ (Age of the Universe [Gyr])}\\
%\multicolumn{3}{ccc}{Hey}\\
%\multicolumn{3}{c}{Hey}\\
%\colhead{$z_f$}&
%\colhead{$z_f$}&
%\colhead{$z_f$}\\
\colhead{$c_{vir}=11.1^a$}&
\colhead{$c_{vir}=6.3^b$}&
\colhead{$c_{vir}=4.6^c$}&
%\colhead{$c_{vir}=4.7^c$}&
%\colhead{Definition}&
\colhead{$c_1^d$}
}
% ~/CLASH/data/a2261/amk/20110601/summary.txt
\startdata
%11.1 & 6.3 & 4.7 & & concentration\\
%2.9 & 1.2 & 0.6 & 3.5\\
%2.3 & 0.9 & 0.4 & 4.1\\
%1.7 & 0.5 & 0.13 & 5.1\\
2.9 (2.2) & 1.2 (5.0) & 0.6 (7.8) & 3.5\\
2.3 (2.8) & 0.9 (6.2) & 0.4 (9.2) & 4.1\\
1.7 (3.8) & 0.5 (8.4) & 0.13 (11.8) & 5.1\\
\vspace{-0.1in}
\enddata
%\tablecomments{Assumes $c_{vir} \approx c_{1} a_{obs} / a_{f}$.}
\tablecomments{Based on $c_{vir} \approx c_{1} (1 + z_{f}) / (1 + z_{obs})$.}
\tablenotetext{a}{\cite{Umetsu09} result.}
\tablenotetext{b}{This work: spherical halo.}
\tablenotetext{c}{This work: elongated halo.}
\tablenotetext{d}{Normalization according to each definition, respectively \cite{Cavaliere11b, Wechsler02, SadehRephaeli08}.}
\end{deluxetable}
%%%%%%%%%%%%%%%%%%%%%%%%%%%%%%%%%

We present results based on these various definitions in Table \ref{tab:obs}.
Regardless of the definition, we note that 
the \cite{Umetsu09} result of $c_{vir} \sim 11$ implies a formation time ($1.7 \lesssim z_f \lesssim 2.9$)
several billion years earlier than our primary result for a spherical halo 
$c_{vir} \sim 6.3$ ($0.5 \lesssim z_f \lesssim 1.2$).
This, in turn, implies a formation time several billion years earlier
than our result for an elongated halo $c_{vir} \sim 4.6$ ($0.13 \lesssim z_f \lesssim 0.6$).
The lone $z_{f} < z_{obs} = 0.225$ result would suggest the cluster has yet to finish ``forming''
according to the \cite{SadehRephaeli08} definition.

Concentration may be the observable most tightly correlated with age for relaxed clusters \citep{WongTaylor11},
but other probes may also be brought to bear.
%But other observational probes may also be brought to bear regarding the cluster formation time.
\cite{Smith10} studied 
BCG morphology, 
luminosity gap $\Delta m_{12}$ between the brightest and second brightest cluster galaxy,
substructure fraction $f_{sub}$,
and cool core strength,
as well as concentrations (as available from X-ray analyses in \citealt{Sanderson09})
in a sample of 59 massive clusters, including A2261.
A2261 was found to be one of four ``fossil clusters'' with a large luminosity gap
$\Delta m_{12} > 2$.
Clusters with $\Delta m_{12} \gtrsim 1$ were found to have 
less substructure,
stronger cool cores,
and
higher mass concentrations,
all likely signatures of earlier formation times without recent major mergers.
In these clusters, the BCG has presumably had time to grow and accrete a significant fraction of
the substructure mass \citep[see also][]{Ascaso11}.
%A measurement of $f_{sub}$ was unavailable from \cite{Smith05}.

%Based on X-ray observations, A2261 is only a borderline cool core cluster.
Based on X-ray observations, A2261 is borderline relaxed (see discussion in \S\ref{sec:c-M})
and a borderline cool core cluster.
Though the temperature profile dips down in the core \citep{Cavagnolo09},
the central entropy floor ($K_0 = 61 \pm 8$ keV cm$^2$) is higher
and the density profile slope ($\alpha \sim -0.7$ at $0.04 ~r_{500}$) shallower
than generally found  ($K_0 < 30$ keV cm$^2$ and $\alpha \lesssim -0.85$)
for cool core clusters \citep{Sanderson09,Smith10}.
There is no obvious star formation visible in the NUV/optical as often found in cool core clusters.
A radio source aligned with the BCG is detected
with $\sim$ 5.3 mJy $\sim 8\times 10^{23}$ W Hz$^{-1}$ at 1.4 GHz
in NVSS \citep{NVSS} 
and 3.39 mJy at 21 cm
in FIRST (\citealt{FIRST}).
All 69 radio-bright ($> 2\times 10^{23}$ W Hz$^{-1}$ at 1.4 GHz) BCGs analyzed by \cite{Sun09} 
were found to be in X-ray cool cores.
%(other tracers?)

Ultimately, analyses of these various observables in CLASH clusters
and in simulated clusters with similar properties
will contribute to significant advancements in our understanding of structure formation and evolution.

\section{Conclusions}
\label{sec:conclusions}

We performed the first robust joint strong and weak lensing analysis of the galaxy cluster A2261. 
We find a halo 
virial mass \Mres\ and concentration \cres\ when assuming a spherical halo.
These tight constraints were enabled through a combination of the 16-band imaging from CLASH 
with multiband wide-field imaging from the Subaru and KPNO telescopes. 
The results show that A2261 is not ``over-concentrated'' as previously found 
but rather is in good agreement with predictions from \LCDM\ N-body simulations.

To explore halo elongation along the line of sight,
we also derived a mass profile based on Chandra X-ray data,
finding it to be $\sim 35\%$ below the lensing mass profile at  $r_{2500}$ ($\sim 500$ kpc). 
This deficit may be explained by an axis ratio of $\sim$ 2:1 outside the inner core 
$r \sim 100$ kpc, corresponding to the visible extent of the BCG.
This elongated mass profile has a lower spherically-defined virial mass
$M_{vir} = 1.65^{+0.16}_{-0.12} \xfmh$
and concentration \cell.  %$c_{vir} = 4.7^{+0.2}_{-0.3}$.
%This should be regarded as a maximal correction which may not be necessary depending on the X-ray results.
Correcting for the lensing effects of massive background structures 
may increase $c_{vir}$ by $\sim 5\%$
and decrease $M_{vir}$ by $\sim 7\%$.
This lower $c_{vir} \sim 4.8$ still agrees with predictions from many simulations
but is lower than predicted by one recent study \citep{Prada11}.

The need to assume halo elongation is critically tied to the reliability of the X-ray mass profile.
Non-thermal pressure support may account partially for the lower X-ray mass.
Published X-ray mass estimates have significant scatter,
including one result in excellent agreement with our spherical lensing mass at $r_{500}$ ($\sim 1.6$ Mpc).

The CLASH survey is providing fundamental and substantial improvements 
in the quantity and quality of observational constraints on cluster dark matter halos. 
Simulations will be tasked with reproducing these empirical results,
contributing significantly to our understanding of structure formation.
Ultimately our results will either confirm \LCDM\ predictions or perhaps yield clues as to the nature of dark energy.

\acknowledgements{
We thank Nobuhiro Okabe for providing his weak lensing shape measurement catalog
for comparison with this work.
We also thank Nobuhiro as well as Mark Voit, Jack Sayers, and Nicole Czakon for useful discussions.
Portions of this collaborative effort were carried out at the Keck Institute for Space Studies,
who we thank for their hospitality.
We thank Perry Berlind and Mike Calkins for efficient operation of the
Hectospec and we thank Susan Tokarz for running the Hectospec data reduction pipeline.

%We are especially grateful to our program coordinator Beth Perrillo 
%for her expert assistance in implementing the \HST\ observations in this program. 
%We thank Jay Anderson and Norman Grogin 
%for providing the \ACS\ CTE and bias striping correction algorithms used in our data pipeline. Finally,
%we are indebted to the hundreds of people 
%who have labored many years to plan, develop, manufacture, install, repair, and calibrate
%the \WFCiii\ and \ACS\ instruments as well as to all those who maintain and otherwise support the operations of the 
%{\em Hubble Space Telescope}.

The CLASH Multi-Cycle Treasury Program is based on observations made with the NASA/ESA Hubble Space Telescope.
The Space Telescope Science Institute is operated by 
the Association of Universities for Research in Astronomy, Inc.~under NASA contract NAS 5-26555.
ACS was developed under NASA contract NAS 5-32864.

This research is supported in part by NASA grant HST-GO-12065.01-A, 
National Science Council of Taiwan grant NSC100-2112-M-001-008-MY3,  % KU 8/11 - 7/14
the Israel Science Foundation, 
%the Baden-Wuerttemberg Foundation, 
the Baden-W\"urttemberg Foundation, 
the German Science Foundation (Transregio TR 33), 
Spanish MICINN grant YA2010-22111-C03-00, 
funding from the Junta de Andaluc\'ia Proyecto de Excelencia NBL2003, 
INAF contracts  ASI-INAF I/009/10/0, ASI-INAF I/023/05/0, ASI-INAF I/088/06/0, PRIN INAF 2009, and PRIN INAF 2010, 
NSF CAREER grant AST-0847157, % Saurabh Jha
the UK's STFC, 
the Royal Society, 
and 
the Wolfson Foundation.
KU acknowledges support from the Academia Sinica Career Development Award.
%AZ acknowledges support by the John Bahcall excellence prize. 
MJG, KJR, AD, and MJK acknowledge partial support from the Smithsonian Institution.
KJR was funded in part by a Cottrell College Science Award from the Research Corporation. 
AD gratefully acknowledges partial support from the INFN grant PD51 and the PRIN-MIUR-2008 grant \verb"2008NR3EBK_003" ``Matter-antimatter asymmetry, dark matter and dark energy in the LHC era''.
LI acknowledges support from a Conicyt FONDAP/BASAL grant.
PR and SS acknowledge support from the DFG cluster of excellence Origin and Structure of the Universe program.
}

{\it Facilities:}
\facility{HST (ACS, WFC3)};
\facility{Subaru (Suprime-Cam)};
\facility{KPNO (Mayall)};
\facility{Chandra (ACIS)};
\facility{MMT (Hectospec)}

\bibliographystyle{astroads}
\bibliography{papersb}

%\appendix
%\section{Virial mass, concentration, and the NFW profile}
%\label{sec:appendix}

\end{document}

%% file: figs/arcs/tabarcs_patched.tex
\begin{deluxetable*}{rccccccccccc}
\tablecaption{\label{tab:arcs}Multiple Images of Galaxies Strongly Lensed by Abell 2261}
\tablehead{
\colhead{}&
\colhead{R.A.}&
\colhead{Decl.}&
\colhead{Magnitude}&
\colhead{Photometric}&
\colhead{Lens Model}\\
\colhead{ID}&
\colhead{(J2000.0)}&
\colhead{(J2000.0)}&
\colhead{F775W (AB mag)}&
\colhead{Redshift$^a$}&
\colhead{Redshift$^b$}
}
\startdata
1a&
17 22 25.452&
+32 08 25.02&
$23.02 \pm 0.04$&
*$4.39^{+0.03}_{-0.04}$&
4.44\\
b&
17 22 27.295&
+32 07 42.58&
$22.82 \pm 0.04$&
$~$$0.48^{+0.03}_{-0.03}$&
"\vspace{0.02in}\\
\hline
2a&
17 22 26.338&
+32 08 20.54&
$26.81 \pm 0.14$&
*$3.89^{+0.12}_{-0.13}$&
3.81\\
b&
17 22 26.133&
+32 08 18.59&
$25.78 \pm 0.16$&
$~$$4.07^{+0.11}_{-0.21}$&
"\\
c&
17 22 25.897&
+32 08 17.16&
$27.04 \pm 0.20$&
$~$$3.74^{+0.14}_{-0.13}$&
"\vspace{0.02in}\\
\hline
3a&
17 22 28.917&
+32 07 55.13&
$24.69 \pm 0.07$&
$~$$0.26^{+2.88}_{-0.01}$&
3.24\\
b&
17 22 28.912&
+32 07 53.96&
"$^{\rm c}$&
"$^{\rm c}$&
"\\
c&
17 22 26.972&
+32 07 38.49&
$26.63 \pm 0.28$&
$~$$2.60^{+0.35}_{-0.77}$&
"\\
d&
17 22 25.391&
+32 08 18.13&
$27.64 \pm 0.28$&
*$3.38^{+0.11}_{-0.21}$&
"\vspace{0.02in}\\
\hline
4a&
17 22 28.569&
+32 08 08.00&
$24.63 \pm 0.04$&
$~$$3.48^{+0.03}_{-0.03}$&
3.31\\
b&
17 22 28.369&
+32 07 37.51&
$24.12 \pm 0.04$&
$~$$3.38^{+0.07}_{-0.07}$&
"\\
c&
17 22 25.217&
+32 08 13.52&
$24.17 \pm 0.05$&
*$3.40^{+0.04}_{-0.04}$&
"\vspace{0.02in}\\
\hline
5a&
17 22 28.589&
+32 08 07.15&
$25.23 \pm 0.09$&
*$3.92^{+0.17}_{-0.25}$&
3.31\\
b&
17 22 28.323&
+32 07 37.38&
$25.04 \pm 0.08$&
$~$$0.29^{+3.36}_{-0.04}$&
"\\
c&
17 22 25.202&
+32 08 13.71&
$24.91 \pm 0.09$&
$~$$3.35^{+0.10}_{-2.85}$&
"\vspace{0.02in}\\
\hline
6a&
17 22 27.069&
+32 08 09.30&
$24.49 \pm 0.07$&
$~$$3.26^{+0.01}_{-0.07}$&
3.09\\
b&
17 22 27.115&
+32 08 01.76&
$21.79 \pm 0.01$&
$~$$0.22^{+0.01}_{-0.03}$&
"\\
c&
17 22 28.615&
+32 07 28.74&
$24.58 \pm 0.07$&
*$3.24^{+0.03}_{-0.03}$&
"\vspace{0.02in}\\
\hline
7a&
17 22 26.599&
+32 07 51.81&
$24.31 \pm 0.07$&
$~$$1.57^{+0.01}_{-0.02}$&
1.74\\
b&
17 22 29.326&
+32 07 59.35&
$23.06 \pm 0.05$&
*$1.54^{+0.01}_{-0.02}$&
"\vspace{0.02in}\\
\hline
8a&
17 22 26.158&
+32 08 25.87&
$25.87 \pm 0.14$&
*$4.92^{+0.13}_{-0.13}$&
4.93\\
b&
17 22 26.737&
+32 07 41.61&
$27.17 \pm 0.17$&
$~$$4.82^{+0.05}_{-0.11}$&
"\vspace{0.02in}\\
\hline
9a&
17 22 25.304&
+32 08 21.19&
$27.11 \pm 0.37$&
*$0.78^{+3.16}_{-0.19}$&
4.46\\
b&
17 22 27.223&
+32 07 39.20&
$26.41 \pm 0.19$&
$~$$4.67^{+0.13}_{-0.15}$&
"\vspace{0.02in}\\
\hline
10a&
17 22 26.379&
+32 08 01.89&
$24.01 \pm 0.05$&
$~$$1.68^{+0.04}_{-0.05}$&
1.67\\
b&
17 22 26.261&
+32 08 01.88&
$24.29 \pm 0.07$&
*$1.79^{+0.01}_{-0.23}$&
"\vspace{0.02in}\\
\hline
11a&
17 22 26.854&
+32 08 06.50&
$22.43 \pm 0.03$&
*$3.88^{+0.04}_{-0.04}$&
3.90\\
b&
17 22 26.875&
+32 08 05.01&
"$^{\rm c}$&
"$^{\rm c}$&
"\vspace{0.02in}\\
\hline
12a&
17 22 26.563&
+32 08 12.81&
$26.42 \pm 0.21$&
$~$$2.42^{+0.05}_{-0.70}$&
2.83\\
b&
17 22 29.045&
+32 07 34.33&
$27.50 \pm 0.15^{\rm b}$&
*$2.43^{+0.34}_{-0.26}$&
"
\\
\vspace{-0.1in}
\enddata
\tablenotetext{a}{The best isolated, least contaminated arcs (marked with *) are chosen to provide the input redshift for each system. These are then optimized by the lens model. Note robust photometric redshifts are not expected for arcs fainter than $\sim$26th magnitude. Uncertainties are formally 95\% C.L. Broad, asymmetric error bars generally indicate a bi-modal solution.}
\tablenotetext{b}{Results from LensPerfect}
\tablenotetext{c}{Continuous arcs were analyzed with a single photometric aperture for more robust photometric redshift estimates.}
\tablenotetext{d}{F814W}
\end{deluxetable*}

%% file: figs/tabz.tex
\begin{deluxetable}{ccccc}
\tablecaption{\label{tab:z}MMT/Hectospec Spectroscopic Redshifts
for Galaxies within the Subaru FOV}
\tablehead{
\colhead{R.A.}&
\colhead{Decl.}&
\colhead{}&
\colhead{Redshift}&
\colhead{Cluster}\\
%\colhead{(J2000.0)}&
%\colhead{(J2000.0)}&
\colhead{(J2000 deg)}&
\colhead{(J2000 deg)}&
%\multicolumn{2}{c}{(J2000 degrees)}&
\colhead{Redshift}&
\colhead{Uncertainty}&
\colhead{Member$^a$}
}
\startdata
260.29303&
32.07600&
0.22096&
0.00017&
0\\
260.30228&
32.26368&
0.38051&
0.00020&
0\\
260.32202&
32.27190&
0.13698&
0.00007&
0\\
260.32373&
32.22027&
0.11367&
0.00009&
0\\
260.34888&
32.04460&
0.15224&
0.00008&
0\\
\vspace{-0.1in}
\enddata
%\tablecomments{Table \ref{tab:z} is published in its entirety in the electronic edition of the Astrophysical Journal.  A portion is shown here for guidance regarding its form and content.}
\tablecomments{This table is published in its entirety in the electronic edition of the Astrophysical Journal.  A portion is shown here for guidance regarding its form and content.}
\tablenotetext{a}{Based on dynamical analysis of the cluster caustics.}
\end{deluxetable}